\documentclass{aa}  
\usepackage{graphicx}
\usepackage{txfonts}
\usepackage{graphicx,epsfig,fancyhdr,rotating,amsmath,natbib}
\usepackage{natbib}
\usepackage{xcolor,ulem}
\usepackage{float}
\usepackage{subfig} 
\usepackage{threeparttable}  
\usepackage{hyperref}  
\usepackage{amsmath}
\usepackage{multicol}
\usepackage{tablefootnote}
\usepackage{setspace}[nodisplayskipstretch]
\usepackage{float}
\usepackage{placeins}
\usepackage{caption,sidecap}
\defcitealias{AstroSpol}{Astrophysical Spectropolarimetry}

\newcommand{\lmax}{\ensuremath{\lambda_{\rm max}}}

\begin{document} 

\setlength{\abovedisplayskip}{6pt}
\setlength{\belowdisplayskip}{6pt}
\setlength{\abovedisplayshortskip}{3pt}
\setlength{\belowdisplayshortskip}{3pt}

   \title{Polarimetric and spectropolarimetric observations with FoReRo2: Instrument overview and standard star monitoring\thanks{Based on data collected with the 2-m RCC telescope at Rozhen National Astronomical Observatory.}}

   \author{Yanko Nikolov\inst{1}\fnmsep\thanks{\email{ynikolov@nao-rozhen.org}}, 
          Galin Borisov\inst{1,2}\fnmsep\thanks{\email{gborisov@nao-rozhen.org}}, Stefano Bagnulo\inst{2}, Plamen Nikolov\inst{1}, Rumen Bogdanovski\inst{1} and Tanyu Bonev\inst{1} }
   \institute{1. Institute of Astronomy and National Astronomical Observatory, Bulgarian Academy of Sciences, 72 Tsarigradsko Chauss\'ee Blvd., BG-1784 Sofia, Bulgaria \\
                2. Armagh Observatory and Planetarium, College Hill, Armagh BT61 9DG, Northern Ireland, UK   
         }
   \date{Received 24 November 2025 / Accepted 01 February 2026}
  \abstract{In this paper, we present a description and characterisation of the 2-Channel Focal Reducer Rozhen (FoReRo2), with an emphasis on its polarimetric and spectropolarimetric observation modes. FoReRo2 is a multimode instrument mounted at the Ritchey–Chr\'etien (RC) focus of the 2-meter Ritchey–Chr\'etien-Coude (RCC) telescope at the Bulgarian National Astronomical Observatory (BNAO) -- Rozhen. The primary targets of this instrument include novae, Be/X-ray binaries, symbiotic stars, asteroids, and comets. 
  
  Standard stars with a high degree of polarisation are an essential part of the polarimetric observations. In our sample, HD 204827 shows short-term variability in the position angle (PA), making it unsuitable for calibration. HD 183143 displays intrinsic polarisation variability, but remains stable in terms of the PA. For the first time, we present a statistical study of the K and \lmax\ parameters of Serkowski's law, including the mean value, standard deviation, and distribution. 
  
  To demonstrate FoReRo2's polarimetric capabilities, we present several examples. 
  The recurrent nova RS Oph shows the variability of Serkowski's K parameter, which is due to dust formation within the first few days after the 2021 outburst. These examples also include the imaging polarimetry of comet C/2019 Y4 (ATLAS) and spectropolarimetric observations of the symbiotic star Z And. 
    }

\titlerunning{Polarimetric observations with FoReRo2}
\authorrunning{Yanko Nikolov et al.}
\maketitle
\titlerunning
\autorrunning
\section{Introduction}
Polarisation is a key characteristic of electromagnetic waves, together with intensity and colour,  opening up a new horizon in the field of object characterisation.
The interest in polarimetric and spectropolarimetric observations has increased in recent years, as reflected in several instrumental papers describing various instruments (e.g., WALOP-North, 1.3-m telescope, Skinakas Observatory, \citealt{2024JATIS..10d4005K}; DIPol-UF, Nordic Optical Telescope, \citealt{2021AJ....161...20P}; CasPol, 2.15-m Jorge Sahade Telescope, Complejo Astron\'omico El Leoncito, Argentina, \citealt{2019JATIS...5b8002S}; DIPOL-1, Sierra Nevada Observatory, 90-cm telescope, \citealt{2024AJ....167..137O}; SPARC4, 1.6-m telescope, Observatório do Pico dos Dias, \citealt{2024BASBr..35...44R}; SPHERE/ZIMPOL, VLT, \citealt{2018A&A...619A...9S}; and ToPol, 1.04-m Omicron (West) C2PU facility, Calern Observatory, France \citealt{ToPol}). We note that the latter has been replaced by a copy of the DIPol-UF instrument used at NOT \citep{2021AJ....161...20P}. This trend is further evidenced by the growing number of observations of standard stars, which play a key role in polarimetric calibration (e.g. \citet{2023A&A...677A.144B, 2024MNRAS.535.1586C, 2024A&A...684A.132M,2020A&A...635A..46P,2020BAAA...61B.183L}). At higher energies, this renewed emphasis is exemplified by the launch of the Imaging X-ray Polarimetry Explorer (IXPE), which has reopened the X-ray polarimetry window \citep{2016SPIE.9905E..17W}.

In this paper, we characterise the polarimetric performance of the 2-Channel-Focal-Reducer Rozhen (hereafter, FoReRo2 or FR2 \citep{2000KFNTS...3...13J}). This instrument is attached to the Ritchey–Chr\'etien (RC) focus of the 2-m telescope of the Bulgarian National Astronomical Observatory (BNAO) -- Rozhen. We use a homogeneous dataset of observations of both zero- and high-polarisation standard stars and present a set of examples of scientific applications in this context. We assess the stability and the accuracy of the instrument over nearly a decade of observations, and provide reliable calibration data that can support future scientific use of FR2. 

In Section~\ref{Sec.2}, we review the FoReRo2 and its observing modes. 
In Section~\ref{Sec.3}, we present the observing strategy and the data reduction steps. Section~\ref{spp.hpol} contains the results of observations of stars with high degrees of polarisation.  Section~\ref{Sec.6} showcases a statistical study of $K$ and \lmax of polarised stars, spectropolarimetric observations of the recurrent nova RS Oph and the symbiotic star Z And, as well as polarimetric observations (imaging) of the comet C/2019 Y4 (ATLAS). These observations serve as examples to demonstrate FoReRo2’s operational flexibility and its capability to deliver high-quality polarimetric and spectropolarimetric data for a diverse range of astronomical targets.
\section{Two-Channel-Focal-Reducer Rozhen} 
\label{Sec.2}
 FoReRo2 was delivered to BNAO based on a contract between the Max-Planck Institute for Solar System Research and the Institute of Astronomy (IA) in 2004. Since that time, the FR2 has been used at the f/8 RC focus of the 2m telescope, where it reduces the original focal ratio to f/2.8, corresponding to an effective focal length of 5.6\,m. 
The FR2 mounted at the telescope is shown in Fig. \ref{fig.2mF2}.
FR2 is a multi-mode instrument offering a wide variety of observing capabilities:
\begin{figure}[htb]
    \begin{center}
      \includegraphics[width=0.30\textwidth, angle=0]{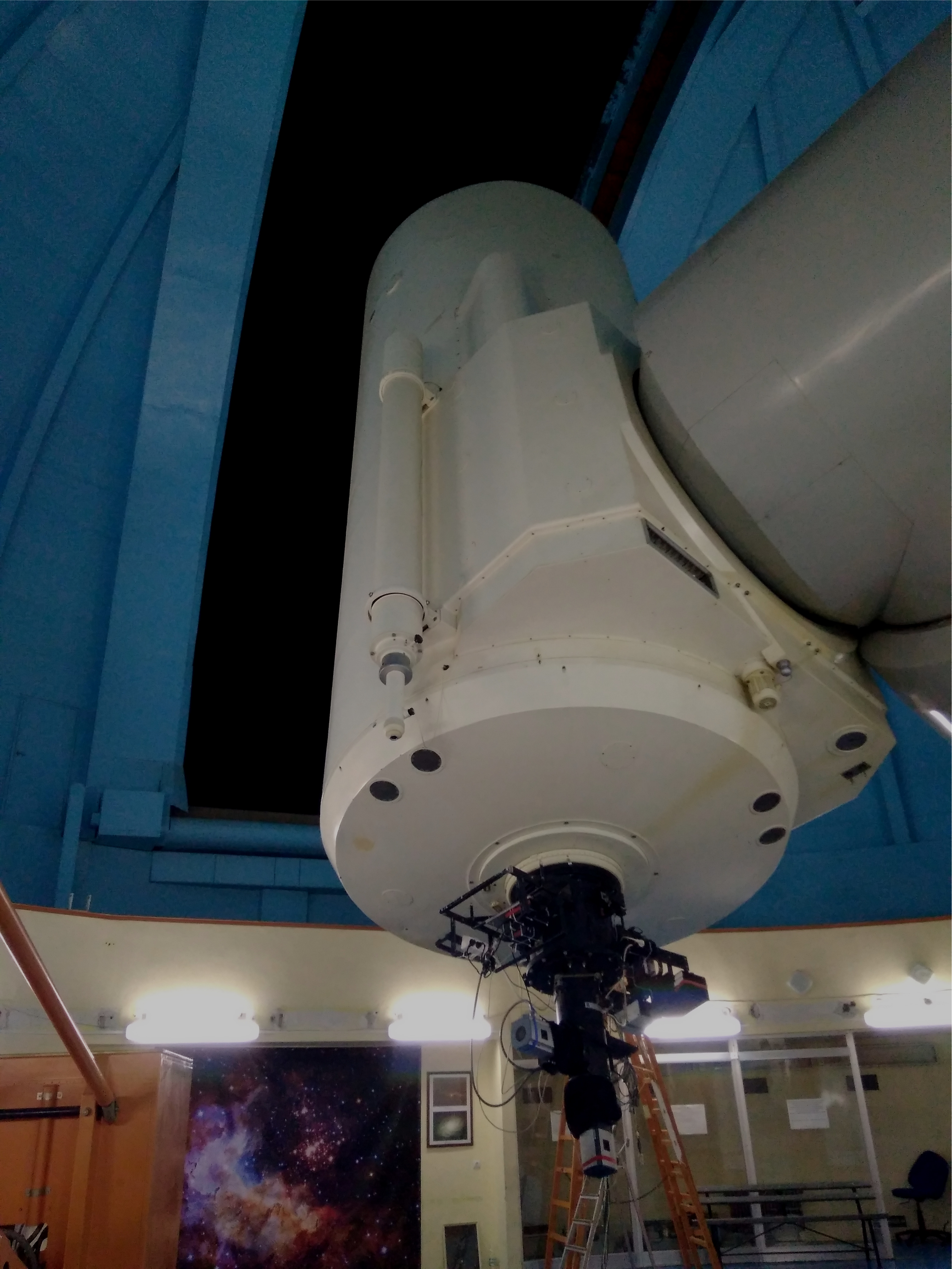}
   \end{center}
         \caption{FoReRo2 attached at the RC focus of the 2m telescope.}
         \label{fig.2mF2}
\end{figure}

 \begin{itemize}
        \item broad-band and narrow-band imaging;
        \item long-slit, low-dispersion (R$\approx1000$) spectroscopy;
        \item Fabry-Perot interferometry;
        \item imaging polarimetry;
        \item low-dispersion spectropolarimetry.         
\end{itemize} 
A detailed description of the instrument is presented in \citet{2000KFNTS...3...13J}.
Since that time, several changes and upgrades of the FR2 have been made. Now only two lenses are available, one in the red and one in the blue channel, colour-corrected in the range of 587--1014\,nm and 365--436\,nm,  respectively. A set of Sloan filters and an H$\alpha$ filter were added to the set of filters listed in \citeauthor{2000KFNTS...3...13J}.

\begin{figure}[htb]
    \begin{center}
      \includegraphics[width=0.30\textwidth, angle=0]{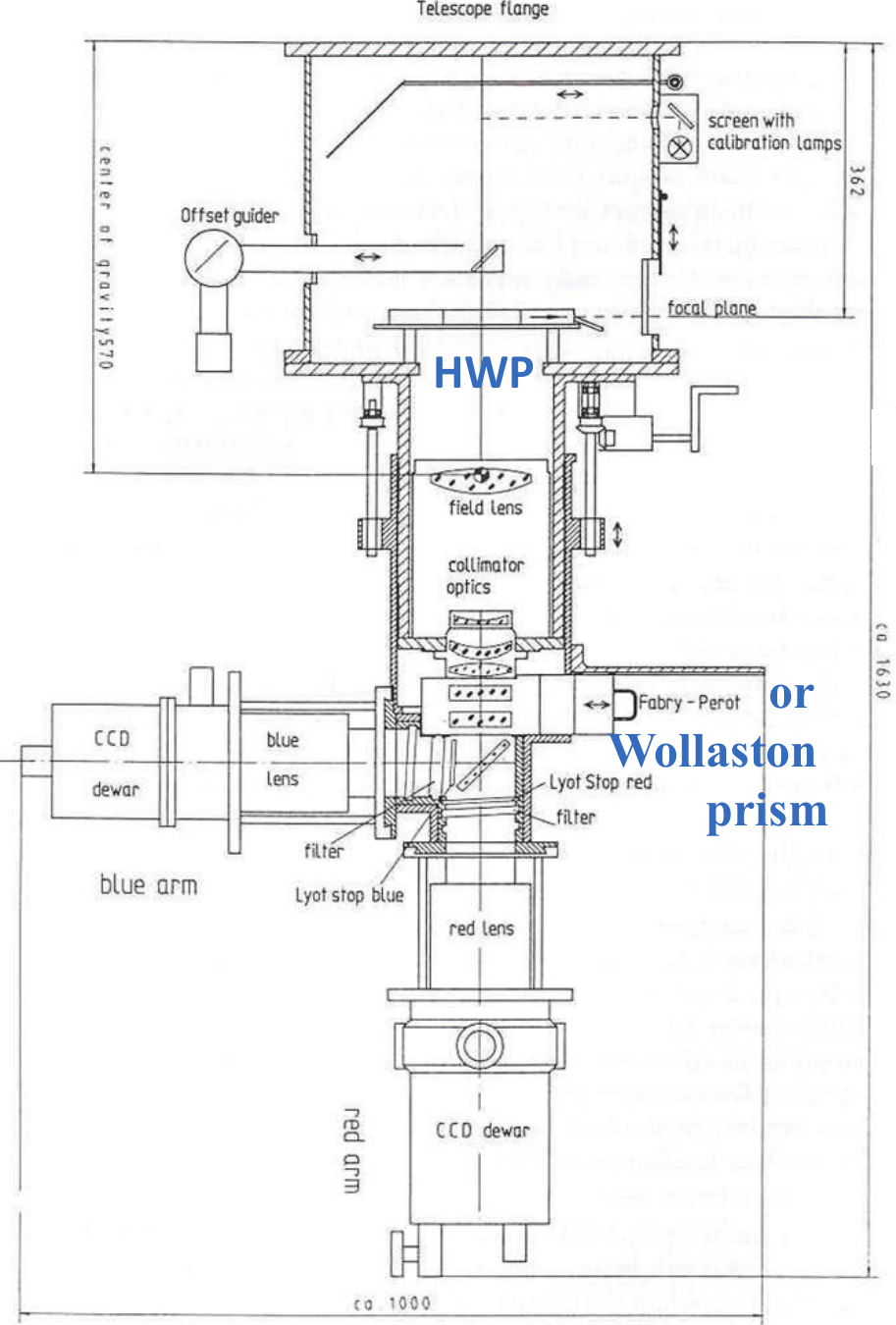}
   \end{center}
         \caption{Drawing of the FoReRo2. The positions of the 'HWP' and 'Wollaston prism' are marked. At that position, the Wollaston is outside of the light beam. }
         \label{fig.FR2}
\end{figure}

One important improvement of the polarimetric mode is a new Wollaston prism, which uses the full aperture of the telescope now. Thus, the previously used quadruple imaging polarimeter, described in  \citet{1996Ap&SS.239..259G}, was replaced with a dual-beam instrument for polarisation measurements. The Wollaston prism is mounted in a special holder, which has the capability of linear translation  (see Fig.~\ref{fig.FR2}). This construction allows a quick switch between the different observing modes. In the case of polarimetric and spectropolarimetric observations, the Wollaston prism is inserted into the parallel light beam, and it is taken out of the beam when switching to other observing modes.     

The Wollaston Quartz prism is placed in the parallel beam, before the colour divider; thus it feeds both channels simultaneously. The split angle of $0.71^\circ$ guarantees separation without overlapping of the ordinary (o) and extraordinary (e) stripes, imaged on the blue and red CCD (see the top panel in Fig.~\ref{fig.a00}).

For several years after the delivery of the FR2 in BNAO,  polarimetric measurements were possible in imaging mode only. In 2011, we modified the instrument in such a way that it can also work in spectropolarimetric mode. This was a turning point which opened the way for low-dispersion spectropolarimetry and the first observation obtained in the spectropolarimetric mode was of comet C/2009 P1 (Garrad)\footnote{http://nao-rozhen.org/observations/fr8.html}.

\begin{figure}[htb]
    \begin{center}
      \includegraphics[width=0.30\textwidth, angle=0]{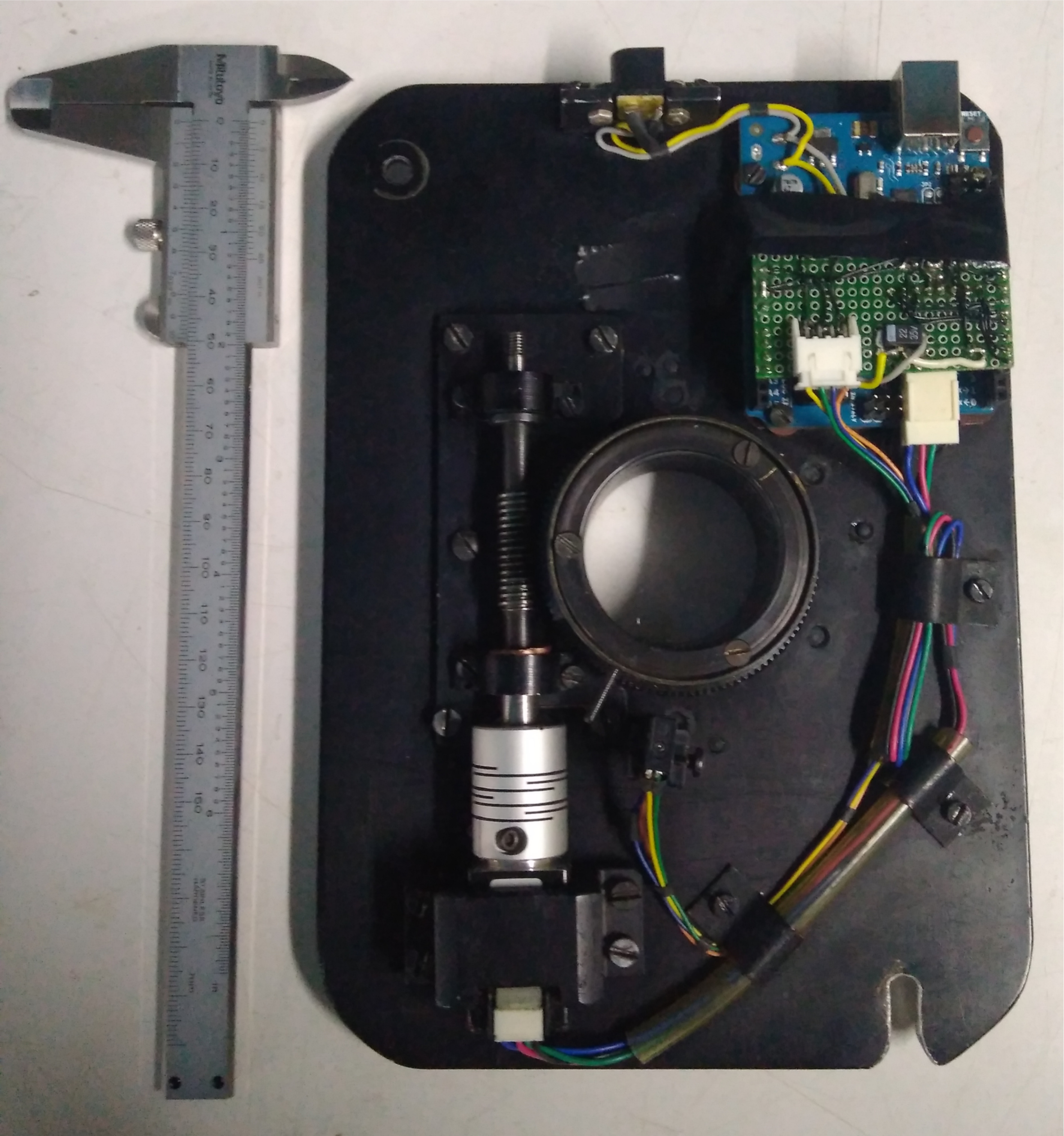}
   \end{center}
         \caption{Half-wave plate and its control unit.}
         \label{fig.hwp}
\end{figure}

For several years, the polarimetric mode was used by rotating the whole FR2 to different angles  to obtain the necessary and sufficient set of observations needed for calculating the Stokes parameters Q and U. In February 2014, the Agreement for collaborative research actions between Armagh Observatory (AO) and Institute of Astronomy and National Astronomical Observatory (IA and NAO) at the Bulgarian Academy of Sciences (BAS) was signed. In the framework of this collaboration, AO delivered to BNAO a $\lambda$/2 retarder plate (half-wave plate (HWP)) for joint polarimetric observations with the 2-m telescope. The design and characteristics of waveplates can be found in \citet{2004JQSRT..88..319S}. The waveplate used with FR2 is a Super-Achromatic True Zero-Order Waveplate~5 - APSAW-5. More details are presented on the webpage of the vendor\footnote{http://astropribor.com/super-achromatic-quarter-and-half-waveplate/}. 
A dedicated, precise mechanical construction and electronic control system for the rotation of the retarder was designed and manufactured in the workshop of the Institute of Astronomy and BNAO. All components are mounted on a plate, which is shown in Fig.~\ref{fig.hwp}. On the bottom side of the plate (not seen in the figure), a gear is mounted that holds the rotation cage (the housing of the HWP has a clear aperture of 40\,mm). A worm drive gear, rotated by a stepping motor, is coupled to the cage. 

The diameter of the $\lambda$/2 retarder is 40\,mm, and it is mounted immediately below the RC focal plane (marked as HWP in Fig~\ref{fig.FR2}). A better solution would be to mount the $\lambda$/2 retarder in the parallel beam, but as the diameter of the latter is 50\,mm, such a solution would cause light losses greater than 35\%. The rotation of the $\lambda$/2 retarder is controlled remotely with a positional accuracy better than $0.1^\circ$. This accuracy was achieved with precise mechanical construction and control electronics consisting of an Arduino board fitted with a stepper motor driver.
%

The Arduino board runs the motor control and precise positioning firmware, which is accessed over a serial interface by a Raspberry Pi single-board computer. The Raspberry Pi computer runs the user software to control the angle of the $\lambda$/2 retarder plate and is accessed remotely over the local area network. To avoid any backslashes in the gears, the firmware is designed in such a way that it always approaches the needed position from one direction only. The absolute angle rotation is achieved using a precise optical null position indicator, which is turned on only during the rotation in order to avoid any light spill during the exposure. The null position indicator is also used for self-calibration.

\begin{figure}[htb]
    \begin{center}
      \includegraphics[width=0.8\columnwidth, angle=0]{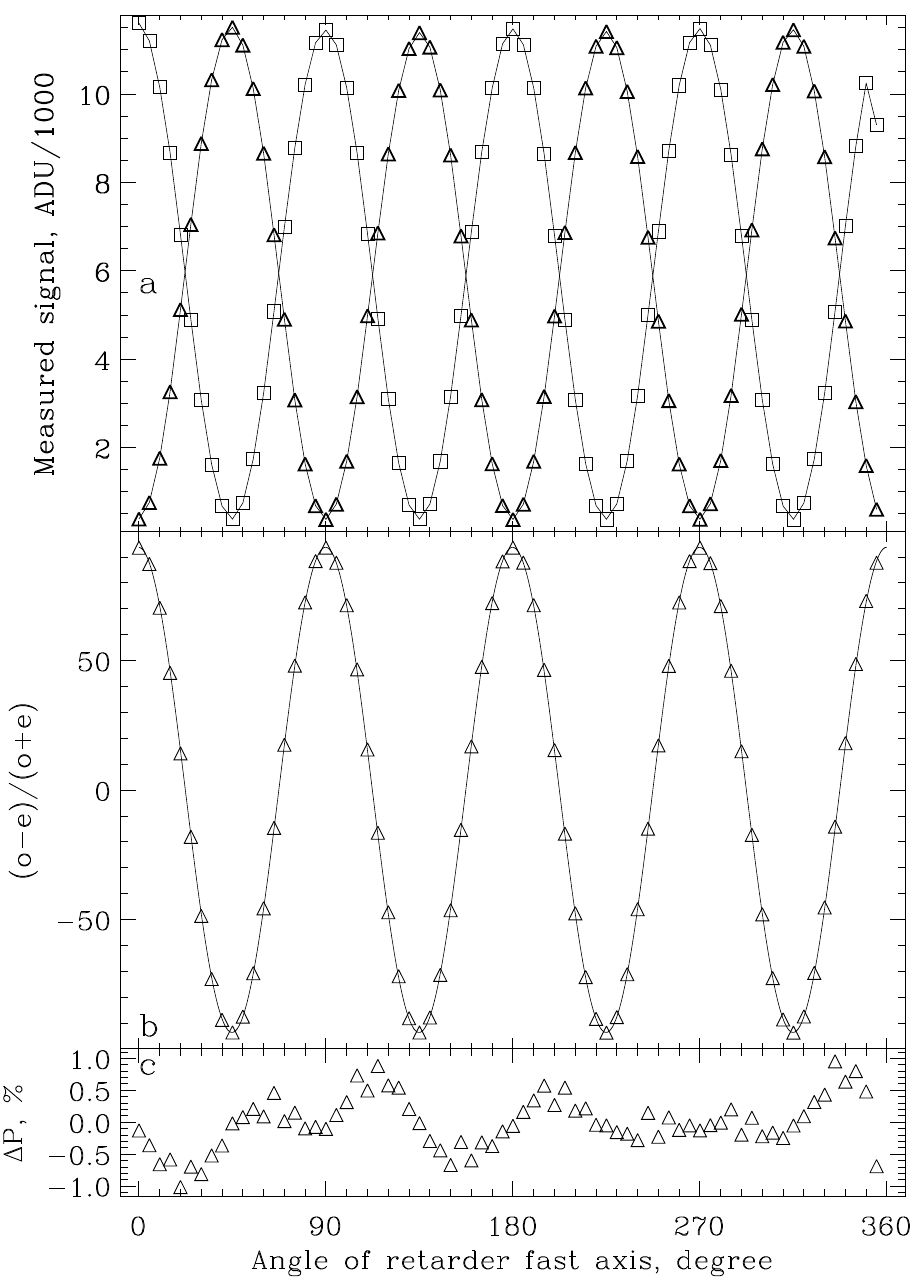}
   \end{center}
         \caption{Fluxes measured over a full rotation cycle of the HWP, 
         Top panel: Fluxes in the o- ($\square$) and e- ($\bigtriangleup$) beams. Middle panel: Measured quantity ($F_{\rm o} - F_{\rm e})/(F_{\rm o} + F_{\rm e})$ and its best-fit. Bottom panel: Residuals. More details are given in the text.}
         \label{fig.zero_point}
\end{figure}

\begin{figure}[htb]
    \begin{center}
      \includegraphics[width=0.40\textwidth, angle=0]{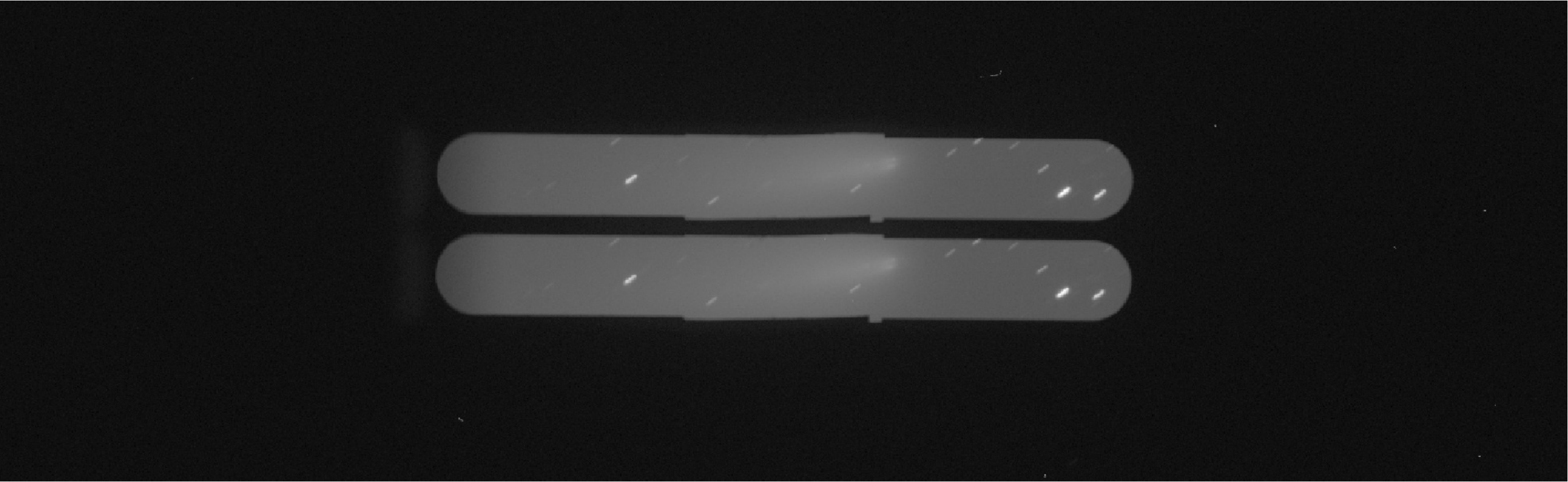}
      \includegraphics[width=0.40\textwidth, angle=0]{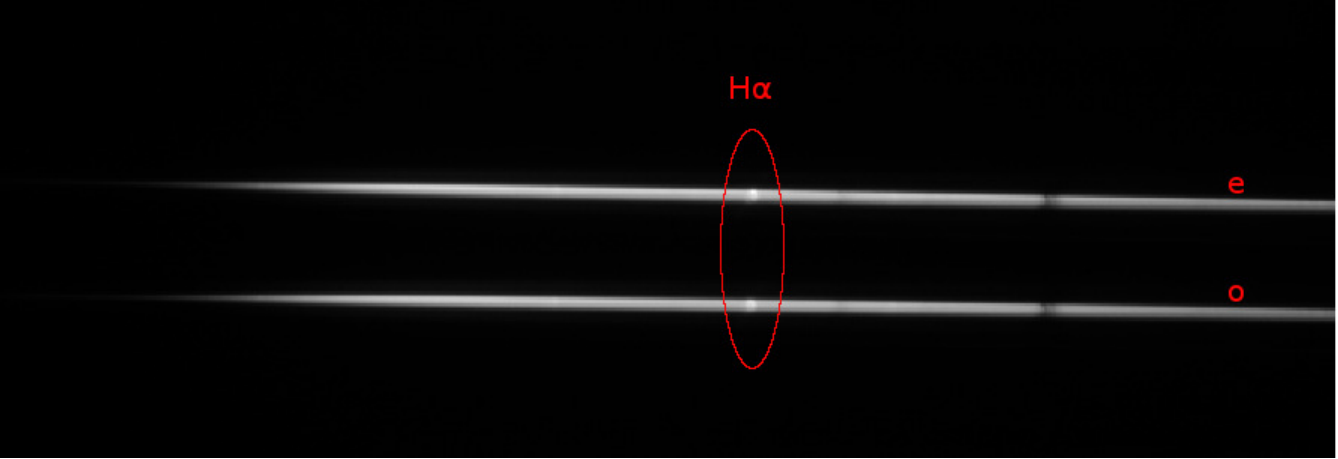}      
   \end{center}
        \caption{Raw frames obtained with FoReRo2 in polarimetric mode. Top panel: Imaging polarimetry of the comet C/2019 Y4 (ATLAS). Bottom panel: Spectropolarimetry of the Be/X-ray binary star X~Per, with the $H_{\alpha}$ emission line visible. Both frames show the o- and e-beam splits by the Wollaston prism.}
         \label{fig.a00}
\end{figure}

The delivered HWP had no mark showing the orientation of its fast axis. Therefore, we performed a laboratory experiment with a twofold aim: (1) to identify the orientation of the fast axis; and (2) to check the mechanical uniformity over a full cycle of rotation. For this experiment, a white screen, inclined at $45^\circ$ with respect to the optical path, was inserted above the RC focal plane. It was illuminated with a tungsten lamp, and the reflected light passed through the HWP. Afterwards, the o- and e-beams were separated by the Wollaston prism. The contrast between the o- and e-beams was increased by putting a Glan prism in the RC focal plane (i.e. between the reflecting screen and the HWP). The Glan prism was rotated manually until the best contrast was achieved (at angle $0^\circ$ of the HWP). Afterwards, the HWP was rotated in steps of $5^\circ$, covering the full range of angles from $0^\circ$ to $360^\circ$. The signal of the o- and e- beams, $F_{\rm o}$ and $F_{\rm e}$, respectively,  was measured in rectangles of equivalent size 56$\times$61 px, positioned around the centres of their respective image stripes. The result of this laboratory experiment is shown in Fig.~\ref{fig.zero_point}. The three panels contain dependencies of the rotation angle of the fast axis of the HWP. In the top panel of this figure, the measured signals of both beams, $F_{\rm o}$ and $F_{\rm e}$, are shown, as squares and triangles, respectively. The middle panel shows the calculated values of the quantity $(F_{\rm o} - F_{\rm e})/(F_{\rm o} + F_{\rm e})$ and its approximation with a $\sin$ function. The bottom panel shows the difference between the measured $(F_{\rm o} - F_{\rm e})/(F_{\rm o} + F_{\rm e})$ values and the fit. These residuals exhibit substantial fluctuations in uniformity during the full cycle of rotation. One relatively uniform region, close to the difference (data-fit) of $0^\circ$, is seen in the range from $220^\circ$ to $310^\circ$. We should note that this laboratory experiment is used with the FR2 only, not with the entire system telescope plus focal reducer.  
%

In 2018, both channels of the FR2 were equipped with new detectors. These are Andor CCD cameras iKon-L 936\footnote{https://andor.oxinst.com/products/ikon-ccd-cameras-for-long-exposure}, with the new dual antireflection deep depletion 'BEX2-DD' chip. Two significant advantages of the new cameras are the thermo-electric cooling and suppression of the fringing in the red and near-IR spectral region, which is especially
useful for the red channel. The cameras comprise 2048$\times$2048 quadratic pixels, each having a size of 13.5\,$\mu$m with a resulting scale of 0.497\,arcsec/px and a FoV of about 17x17\,arcmin (in imaging mode).

For spectropolarimetric observations, we use a grism with 300 lines $mm^{-1}$ in the red channel. Two slit widths are available: 110\,$\mu$m and 200\,$\mu$m. Using the 110\,$\mu$m slit yields a resolving power of $R\approx1000$ at $\lambda=6562\,\AA$, with a reciprocal spectral dispersion of {3~$\AA$~pixel$^{-1}$}. Over the past eight years (after re-coating the 2 m mirror) to obtain more accurate results, we only used slitless spectropolarimetry (see Section~\ref{slit-vs-slitless} for details). In the slitless mode, the resolving power depends on the seeing.

Raw images of the polarimetric (top: comet C/2019 Y4 (ATLAS)) and spectropolarimetric (bottom: Be/X-ray binary star X Per) mode of observations are presented in Fig.~\ref{fig.a00}. For the polarimetric mode of observations, a mask of size 35\,mm $\times$ 4\,mm is used. For slitless spectropolarimetric observations, a mask of size 4\,mm$ \times$ 3\,mm was used. Two spectra of the Be/X-ray binary star X Per are presented in Fig.~\ref{fig.a00}, where the $H_{\alpha}$ emission line is visible. The spectra correspond to the o- and e- beam, respectively. 
  %
%
\section{Observing strategy and data reduction}
\label{Sec.3}
\subsection{Observations and basic reduction steps}
We obtained polarised spectra and polarised images at eight HWP angles $22.5^\circ$ apart. Standard stars with zero degrees of polarisation were used to correct the instrumental polarisation. 
Standard stars with a high degree of polarisation were used to correct the PA.  
Polarisation standards were observed with the same instrumental setup as astronomical objects on the same nights. 

The spectra are reduced using IRAF \citep{1993ASPC...52..173T} in the standard way, including bias removal and wavelength calibration. For the spectropolarimetric data reduction, our research group created dedicated IRAF scripts.
For better comparison with catalogue values in the V filter, we used the extracted $P_V$ and $\theta_{V}$ from spectra using the IRAF SBAND procedure. 

The process of spectropolarimetric data reduction includes the following: bias subtraction, extracting 1D spectra, wavelength calibration, beam-swapping technique, and corrections for instrumental polarisation, chromatism of the HWP, and the PA. The beam-swapping technique (BST) \citep{2009PASP..121..993B} was at the core of the sequence of reduction steps. We used it to minimise the instrumental polarisation. We tested spectropolarimetric data reduction with and without the use of flat fields. 
Our tests show that in both cases, with and without flat fields, the results obtained are indistinguishable within the precision of our data. This result is in agreement with the result obtained by other authors \citep{2019JATIS...5b8002S}.
\subsection{Beam-swapping technique}
A beam-swapping technique is used for polarimetric data processing. 
The values of F$(\lambda)_{\rm o}$ and $F(\lambda)_{\rm e}$, where $F(\lambda)_{\rm o}$ and F$(\lambda)_{\rm e}$ are the fluxes of the o- and e- beams, respectively, obtained for the different HWP angles. We can define 
$\mathcal{D}_{\alpha}=\left(\frac{F(\lambda)_{\rm o} - F(\lambda)_{\rm e}}{F(\lambda)_{\rm o} + F(\lambda)_{\rm e}}\right)_{\alpha}$, where $\alpha$ corresponds to the HWP angle. The following formulae are used to calculate the Stokes parameters \citep{2009PASP..121..993B}:
\begin{eqnarray}
\begin{aligned}
P_Q(\lambda)= \frac{Q(\lambda)}{I(\lambda)} =\frac{1}{4} 
\left(\mathcal{D}_{0^\circ} - \mathcal{D}_{45^\circ} + \mathcal{D}_{90^\circ} - \mathcal{D}_{135^\circ}\right),
\end{aligned}
\end{eqnarray}
\begin{eqnarray}
\begin{aligned}
P_U(\lambda) = \frac{U(\lambda)}{I(\lambda)} =\frac{1}{4}
\left(\mathcal{D}_{22.5^\circ} - \mathcal{D}_{67.5^\circ} + \mathcal{D}_{112.5^\circ} - \mathcal{D}_{157.5^\circ}\right).
\end{aligned}
\end{eqnarray}

\begin{figure}[htb]
    \begin{center}
      \includegraphics[width=0.49\textwidth, angle=0]{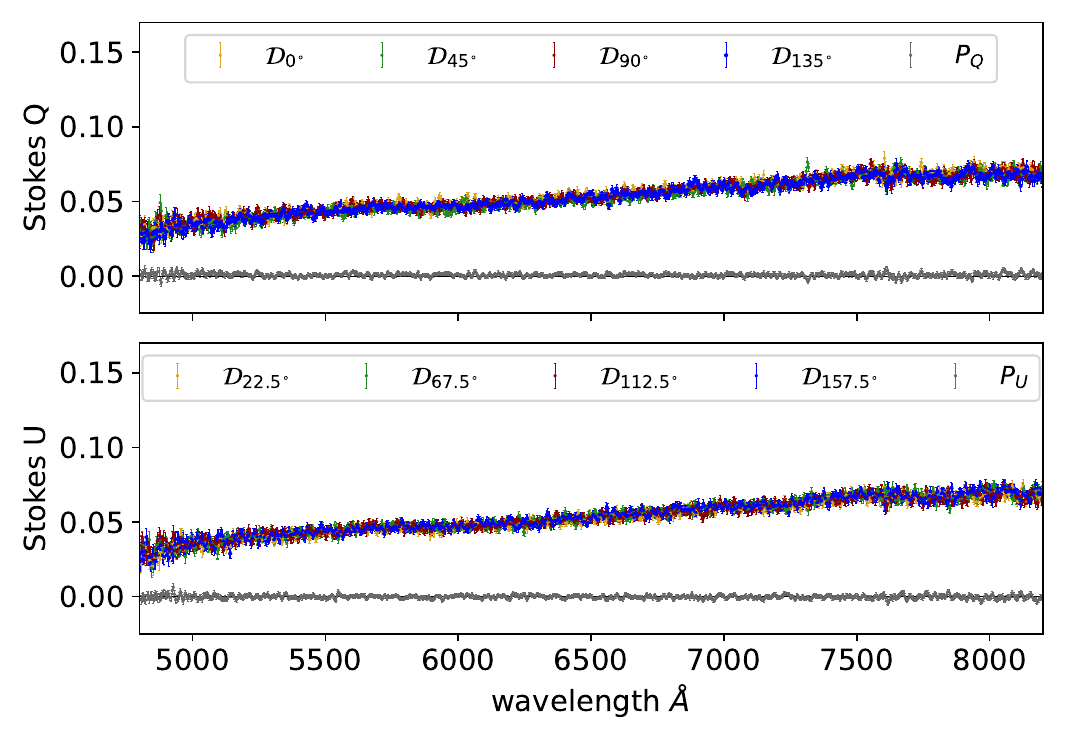}
   \end{center}
         \caption{Stokes Q and U parameters of the unpolarised standard star HD 154982 after applying the beam-swapping technique.}
         \label{fig.bst}
\end{figure}

Figure~\ref{fig.bst} presents the Stokes Q and U parameters of the unpolarised standard star HD 154892 after applying the beam-swapping technique. As Fig.~\ref{fig.bst} demonstrates, after applying the BST, the observed polarisation of the unpolarised standard star is close to zero.
%
%
\subsection{Instrumental polarisation}
Regular spectropolarimetric observations were initiated after the installation of the HWP in November 2015. The BST can reduce instrumental polarisation caused by the optical components located between the HWP and the CCD cameras. The residual instrumental polarisation is introduced by the optical elements before the HWP and the combination of the reflective coating of the primary and secondary mirrors. This residual instrumental polarisation was corrected using standard stars with zero degree of polarisation.
We corrected the Stokes parameters $q_{\rm obs}(\lambda)$ and $u_{\rm obs}(\lambda)$ of the objects for instrumental polarisation using the Stokes parameters $q_{\rm zpol}(\lambda)$ and $u_{\rm zpol}(\lambda)$ of standard stars with zero degree of polarisation using the following equations: 
\begin{eqnarray}
q_{\rm corr}(\lambda)=q_{\rm obs}(\lambda) - q_{\rm zpol}(\lambda), \\
u_{\rm corr}(\lambda)=u_{\rm obs}(\lambda) - u_{\rm zpol}(\lambda),
\end{eqnarray}
where $q_{\rm corr}(\lambda)$ and $u_{\rm corr}(\lambda)$ represent both Stokes parameters corrected for instrumental polarisation. 

\begin{figure}[htb]
    \begin{center}
      \includegraphics[width=\columnwidth, angle=0]{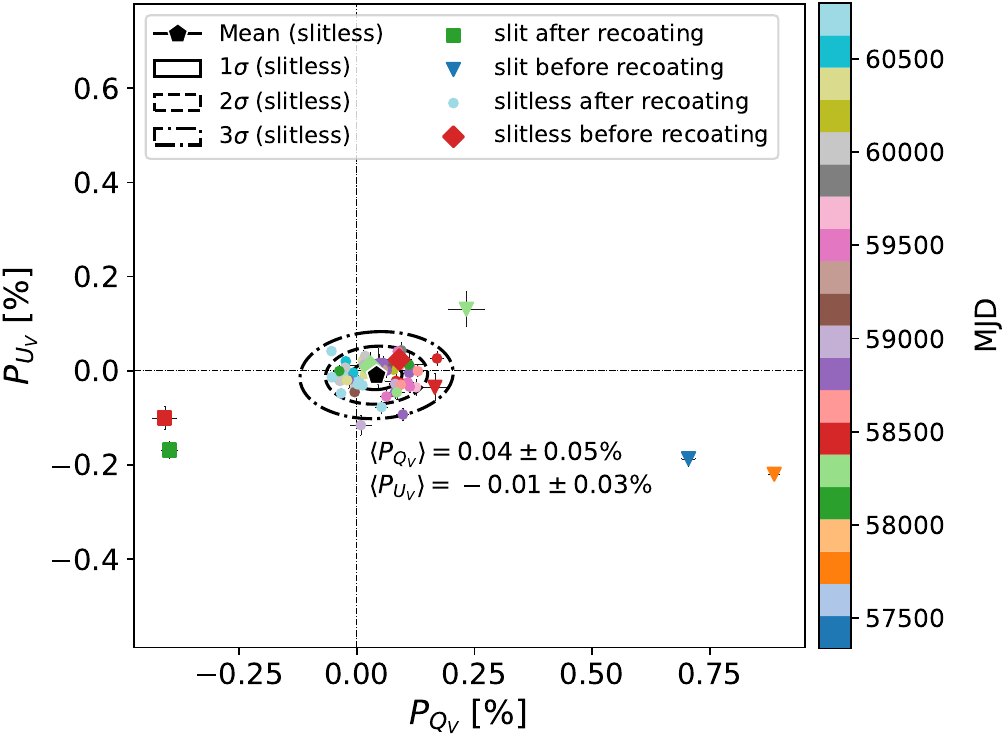}
   \end{center}
         \caption{Q-U diagram of observed zero polarisation standard stars in synthetic V filter before and after the re-aluminisation of the 2-m mirror. The colour of each symbol represents the modified Julian date (MJD).}
         \label{fig.quzpol}
\end{figure}

The 2m mirror's aluminium reflective coating has to be renewed from time to time. The last two aluminium reflective coatings occurred in 2008 and 2017. Figure~\ref{fig.quzpol} shows the Q-U diagram of zero polarisation standard stars in a synthetic V filter before and after the re-coating of the 2m mirror. A synthetic V filter represents the standard Johnson V-band transmission curve and is used in synthetic photometry to simulate photometric measurements from spectral data. When applied (e.g., via IRAF's SBAND task), it multiplies the spectrum by the filter transmission curve and integrates the result to estimate the flux or magnitude as if observed through a real V-band filter. This allows for a direct comparison between the spectro-polarimetric and polarimetric data. 

%

In Figure~\ref{fig.quzpol}, different symbols indicate the spectropolarimetric observation modes (with and without slit, before and after the re-coating of the 2-m mirror), while the colour represents the modified Julian date (MJD). When a slit is used (triangles and squares), the instrumental polarisation is high and, before the mirror re-coating, also highly variable (triangles). In the slitless mode, no significant difference is observed before and after the re-coating. The mean Stokes parameters in the synthetic V filter for the slitless mode are $P_{Q_{V}}=0.04\pm0.05(\%)$ and $P_{U_{V}}=-0.01\pm0.03(\%)$, marked with the black pentagon. The ellipses represent the 1-, 2-, and 3-$\sigma$ standard deviations. The diagram also shows a shift of the slitless-mode points (a decrease in Stokes Q), with the latest data lying to the left of the black pentagon, possibly due to changes in instrumental polarisation and the selection of more accurate zero-polarisation standard stars.

\begin{figure}[h]
    \begin{center}
      \includegraphics[width=0.49\textwidth, angle=0]{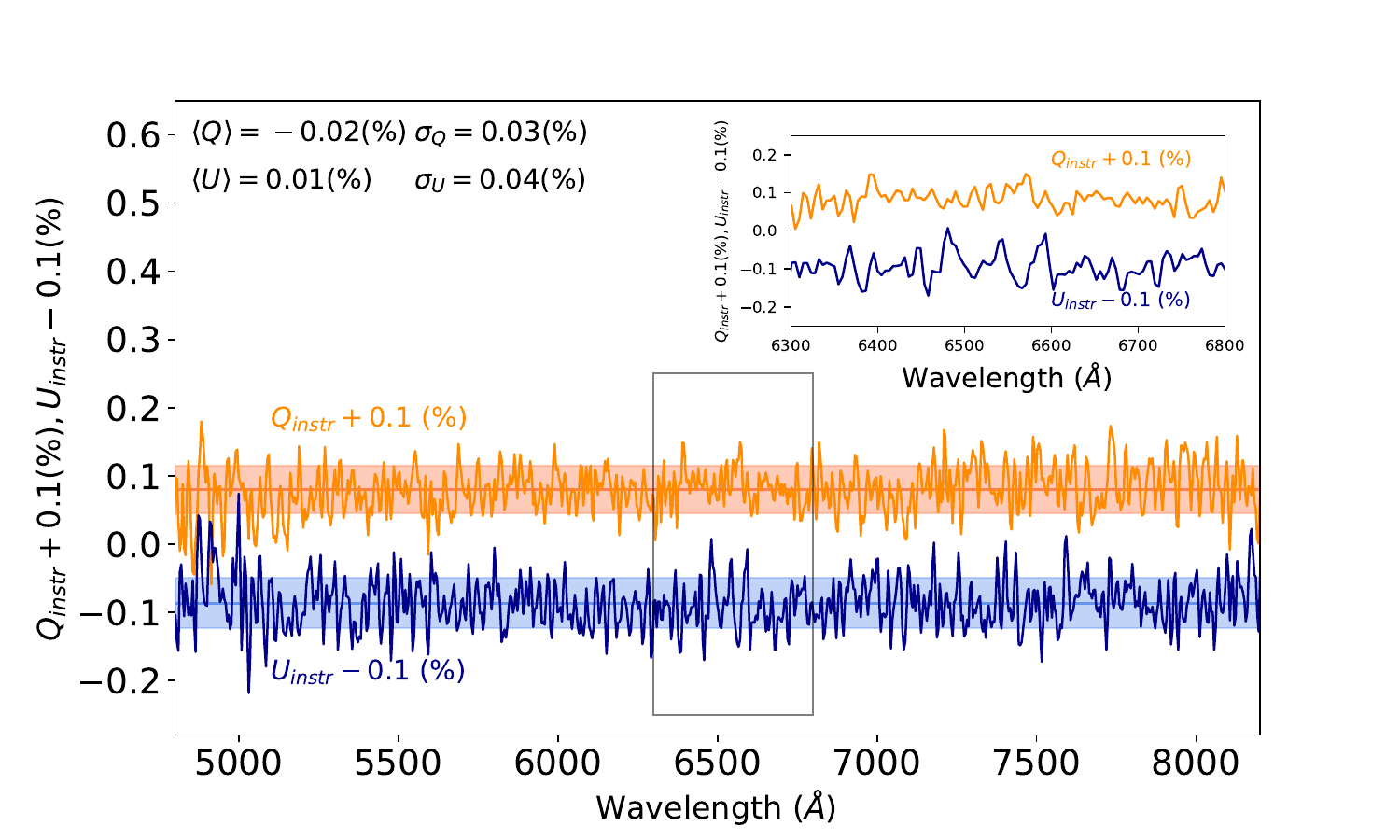}
   \end{center}
         \caption{Instrumental polarisation. The data present an unpolarised standard star, HD 191195. In the observed Stokes Q and U, we do not see any significant (>0.1~\%) ripple pattern.}
         \label{fig.hd191195}
\end{figure}

Figure~\ref{fig.hd191195} shows the polarised spectra of the unpolarised standard star HD 191195 over the wavelength range 4800–8200 \AA. No wavelength dependence is observed in the Stokes parameters. The widths of the orange and blue bands indicate the standard deviation from the mean values, calculated over the entire wavelength range. Multiple reflections within the wave plate (e.g., \citealt{2003SPIE.4843..437I, 2006AJ....132..433K}) can produce the so-called ripple effect; however, no significant (>0.1\%) ripple pattern is detected in the observed Stokes Q and U.
%
\subsection{Slit versus slitless spectropolarimetry}\label{slit-vs-slitless}
\begin{figure}[htb]
    \begin{center}
      \includegraphics[width=0.99\columnwidth, angle=0]{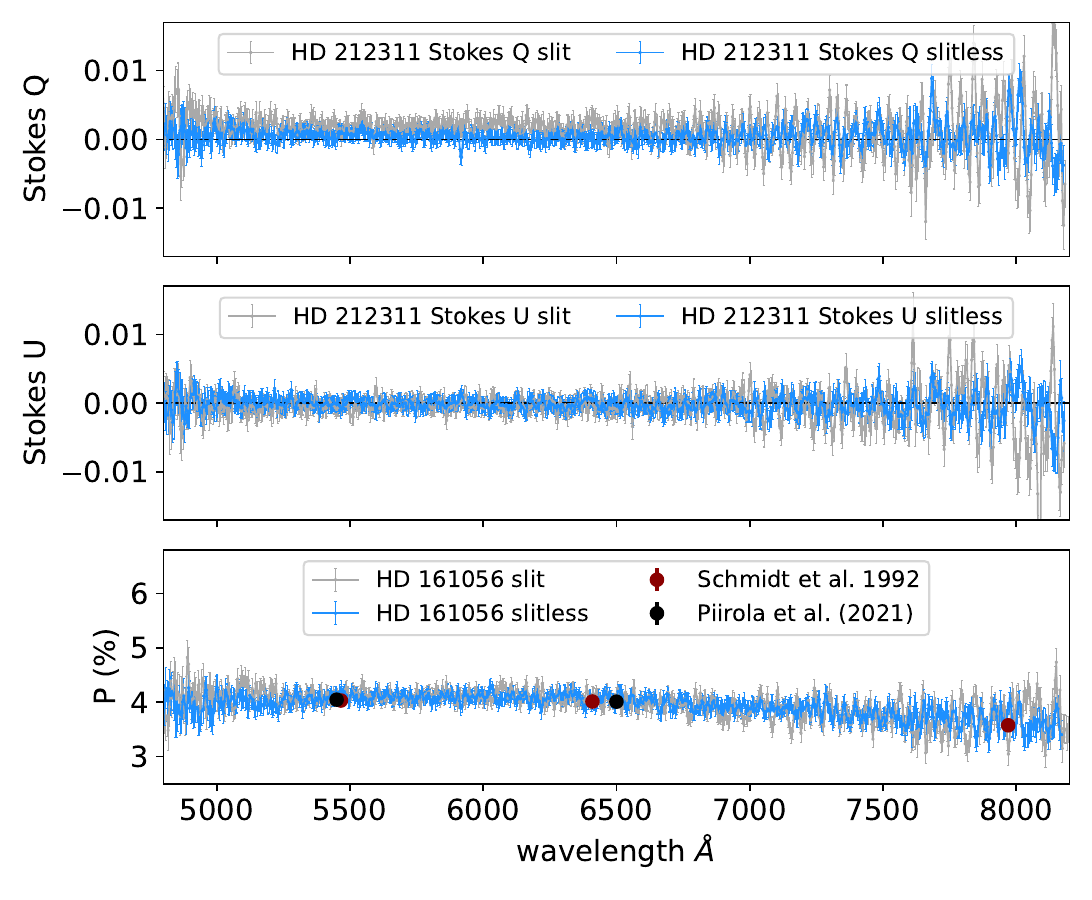}
   \end{center}
         \caption{Slit vs. slitless spectropolarimetry before the re-coating of the 2m mirror.}
         \label{fig.sl.before}
\end{figure}

\begin{figure}[htb]
    \begin{center}
      \includegraphics[width=0.99\columnwidth, angle=0]{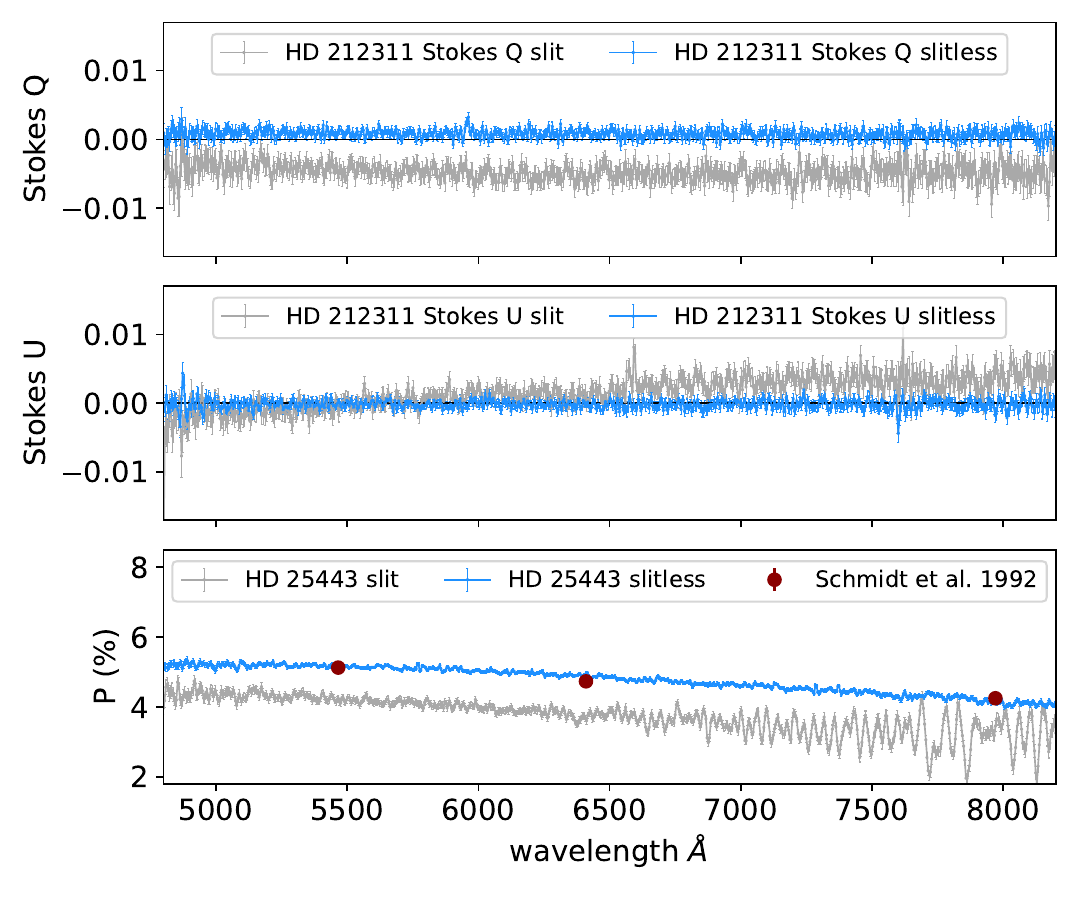}
   \end{center}
         \caption{Slit vs. slitless spectropolarimetry after the re-coating of the 2m mirror.}
         \label{fig.sl.after}
\end{figure}

Two types of observations of zero- and high-polarisation standards were performed:\ slit and slitless spectropolarimetric observations before and after the re-coating of the 2m mirror, respectively. The results of the slit and slitless spectropolarimetric observations before and
after the re-coating of the 2m mirror are presented in Figs.~\ref{fig.sl.before}
and~\ref{fig.sl.after}, respectively.

Prior to the re-coating of the 2m mirror, the results of slit and slitless spectropolarimetric 
observations are practically identical (Fig.~\ref{fig.sl.before}). 
After the re-coating of the 2m mirror, the results from the slit and slitless spectropolarimetry 
observations exhibit clear differences (Fig.~\ref{fig.sl.after}).

It is known that when the width of a slit is comparable to the wavelength of the incident light, the slit acts as a partial polariser. Even though in our case the slit width is around 150 times the H$_\alpha$ wavelength based on the theory by \citet{Slater42}, rooted in microwave wave-guide theory (which also includes the depth of the slit and its material conductivity), we can show that the instrumental polarisation caused only by the slit is in the order of a few per cent. This is aptly explained by \citet{AstroSpol-Keller}\footnote{See \textit{\citetalias{AstroSpol}}, ed. by \citeauthor{AstroSpol}.}.

In our case, this addition to the instrumental polarisation is most likely due to the combination of the explained slit effect, as well as a small misalignment of the 2m mirror relative to the optical axis, which leads to a slightly deformed star PSF (coma), as well as the narrow metal slit, which cuts the star PSF asymmetrically. This is exactly what you can see on Figures~\ref{fig.quzpol} and \ref{fig.sl.after}.
Before and after the mirror re-aluminisation, the Stokes Q has a different value, which is in the order of 1\%.
Figure~\ref{fig.sl.after} even shows how the Stokes U is affected by the instrumental polarisation caused by the coma-slit combination and its wavelength dependence. The results of slitless spectropolarimetric observations are presented in Section~\ref{spp.hpol}.
It is only slitless spectropolarimetric observations that offer the ability to properly correct for the instrumental polarisation. 
%
%
%
\subsection{Correction for the chromatism (wavelength dependence) of the half-wave plate}
\begin{figure}[htb]
    \begin{center}
      \includegraphics[width=0.50\textwidth, angle=0]{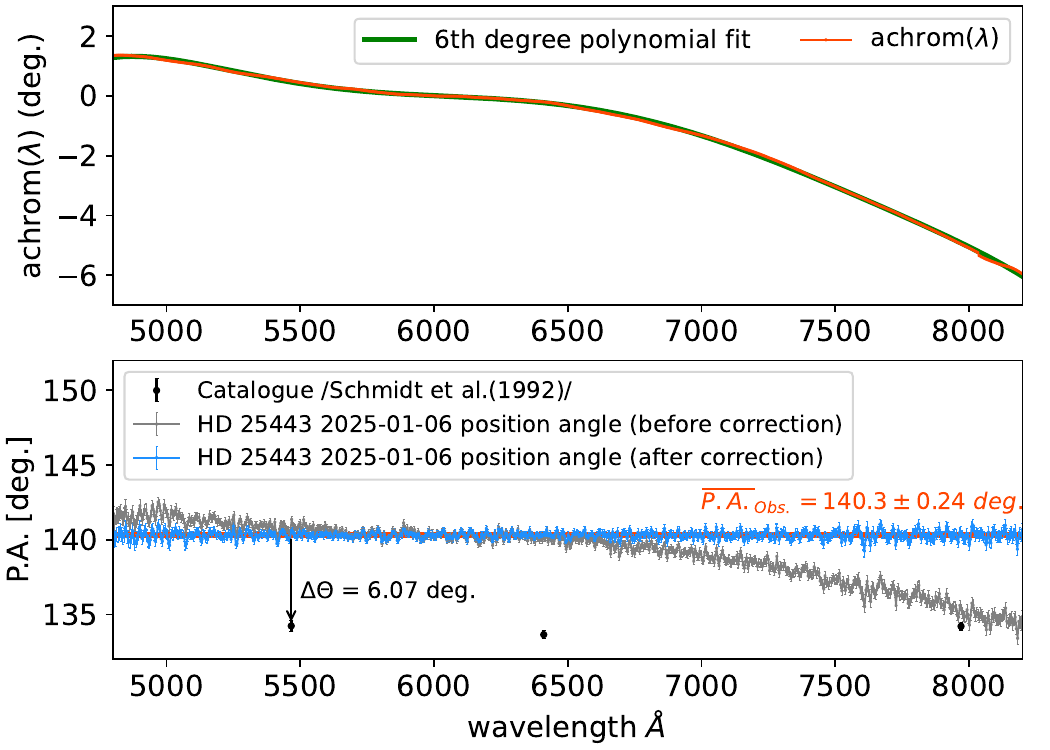}
   \end{center}
         \caption{Correction for the chromatism of the retarder waveplate.}
         
         \label{fig.ach}
\end{figure}

For correction of the chromatism of the HWP, we used the wavelength dependence function $(ach(\lambda))$, which is represented in Fig.~\ref{fig.ach} (top panel). This function was derived from the wavelength dependence of the observed PA of the high polarisation standard star HD 161056. The achromatic function is well described by a sixth-degree polynomial fit:
\begin{equation}
{\arraycolsep=1.4pt
\begin{array}{rcl}ach(\lambda) & =\; & -9690.91 + 9.15\,\lambda - 3.57 \times 10^{-3} \lambda^2 \\
              & & + 7.35 \times 10^{-7} \lambda^3 - 8.43 \times 10^{-11} \lambda^4 +\\
              & & + 5.12 \times 10^{-15} \lambda^5 - 1.28 \times 10^{-19} \lambda^6.
\end{array}}
\end{equation}

We used the following equations to correct for the chromatism (wavelength dependence) of the HWP. 
\begin{equation}
{\arraycolsep=1.4pt
\begin{array}{rcl}
q_{ach}(\lambda) &=& \phantom{-}q_{\rm corr}(\lambda) \, \cos\!\left( 2 \, ach(\lambda) \right) 
+ u_{\rm corr}(\lambda) \, \sin\!\left( 2 \, ach(\lambda) \right),\\
u_{ach}(\lambda) &=& -q_{\rm corr}(\lambda) \, \sin\!\left( 2 \, ach(\lambda) \right) 
+ u_{\rm corr}(\lambda) \, \cos\!\left( 2 \, ach(\lambda) \right),
\end{array}}
\end{equation}
where 
$q_{ach}(\lambda)$ and $u_{ach}(\lambda)$ represents Stokes parameter corrected for the chromatism. We used these values of Stokes parameters to calculate $\theta_{ach}(\lambda)$. The results of this correction are represented in Fig.~\ref{fig.ach} (bottom panel), where gray data represent the PA before chromatism correction, while the blue data show it after the correction.
%
%

\subsection{Correcting the position angle}
\begin{figure}[htb]
    \begin{center}
      \includegraphics[width=0.53\textwidth, angle=0]{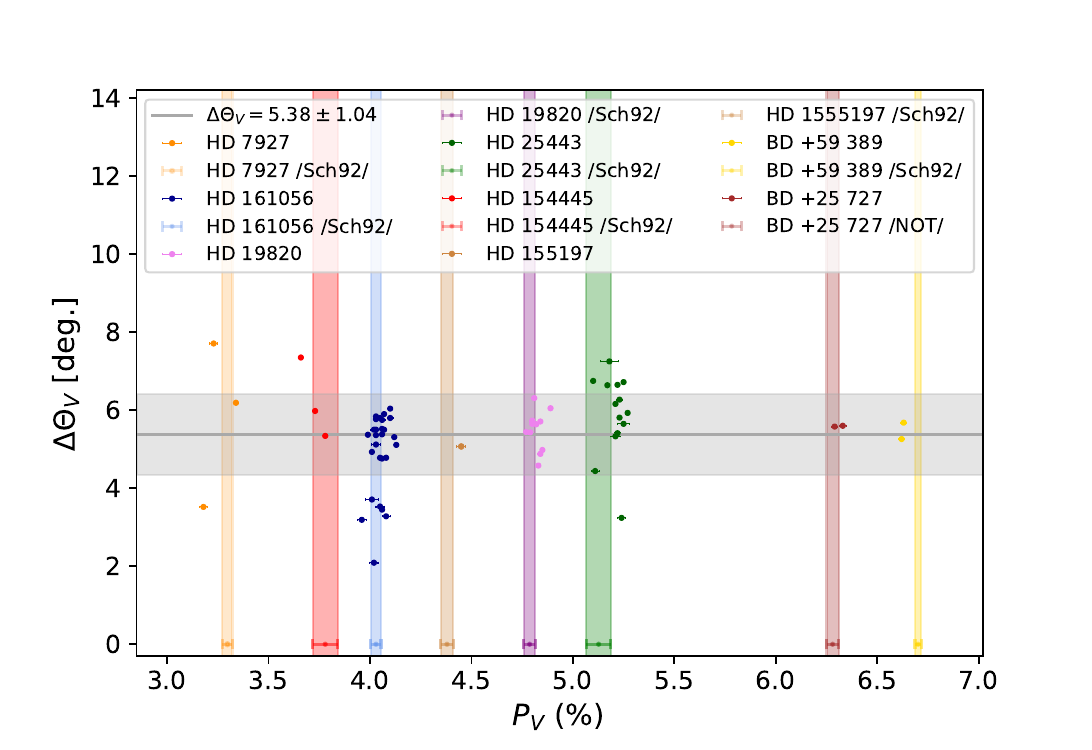}
   \end{center}
         \caption{Observed degree of polarisation in the synthetic V band and the corresponding PA correction ($\Delta\theta$) for high-polarisation standard stars. Vertical coloured bands indicate the catalogue degree of polarisation with its uncertainty (taken from \citet{1992AJ....104.1563S}) and Nordic Optical Telescope (NOT). The horizontal grey line marks the average $\Delta\theta$.}
         \label{fig.hpol.theta}
\end{figure}

The final correction applied during data reduction is the correction for the PA. Standard stars with a high degree of polarisation were used to correct the PA. Here, $\Delta\theta$ represents the difference between the observed values in the synthetic V filter and the catalogue value. The correction follows the equation, 
\begin{equation}
\theta(\lambda)=\theta_{ach}(\lambda) - \Delta\theta ,
\label{delta.theta3}
\end{equation} 
where $\Delta\theta$ represents the difference between the observed and catalogue values of the PA. The average value of $\Delta\theta = 5.38^\circ \pm 1^\circ.04$. 
The RMS of $\Delta\theta$ is $1^\circ.04$, which indicates the instrument's stability with respect to the PA correction. 
In Fig. \ref{fig.hpol.theta} we present the observed degree of polarisation in the synthetic V band and $\Delta\theta$. The vertical columns represent catalogue values of the degree of polarisation in the V band and the width of the column corresponds to the uncertainties in the catalogue values.  The horizontal grey line represents the average value of $\Delta\theta$. 
This offset in the PA depends on the instrument alignment as well as asymmetric ageing of the mirror coating from season to season. For this reason, it is necessary to determine these PA offsets for each observation night or each observation season.
%
%

\subsection{Selection of aperture for imaging polarimetry.}
\label{select.aperture}
\begin{figure}[htb]
    \begin{center}
      \includegraphics[width=0.6\columnwidth, angle=90]{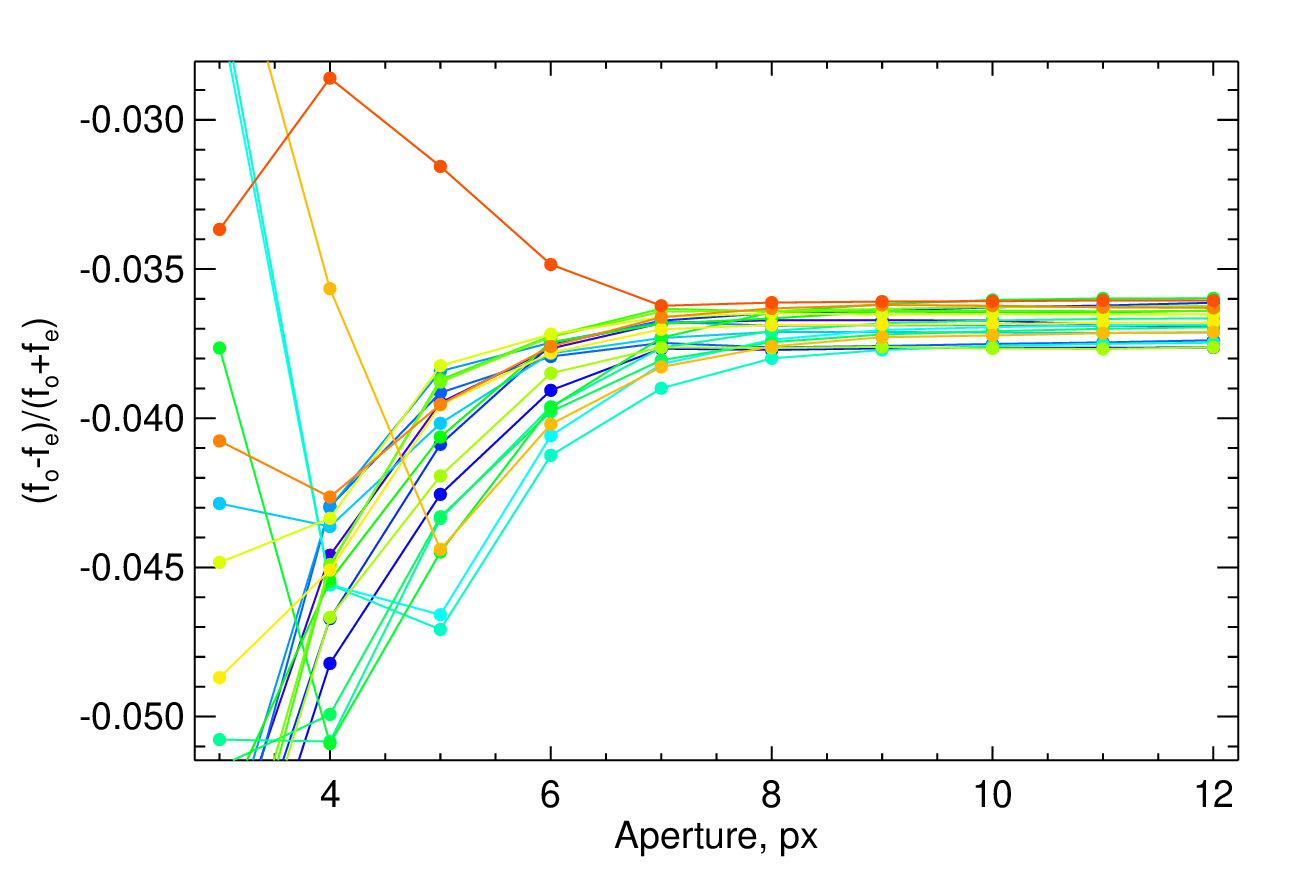}
   \end{center}
         \caption{Example of the best aperture selection for imaging polarimetry for one of the unpolarised standard stars (here, we chose 9\,px). }
         \label{FIG:aper}
\end{figure}

\begin{figure*}[htb]
    \begin{center}
      \includegraphics[width=\columnwidth, angle=0]{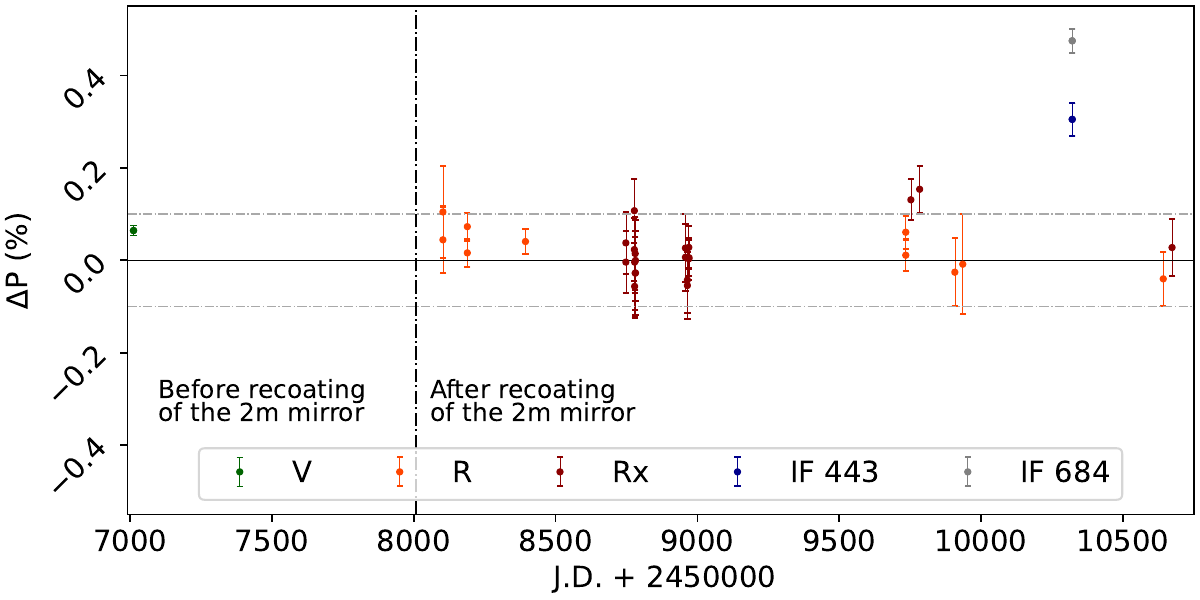}
      \includegraphics[width=\columnwidth, angle=0]{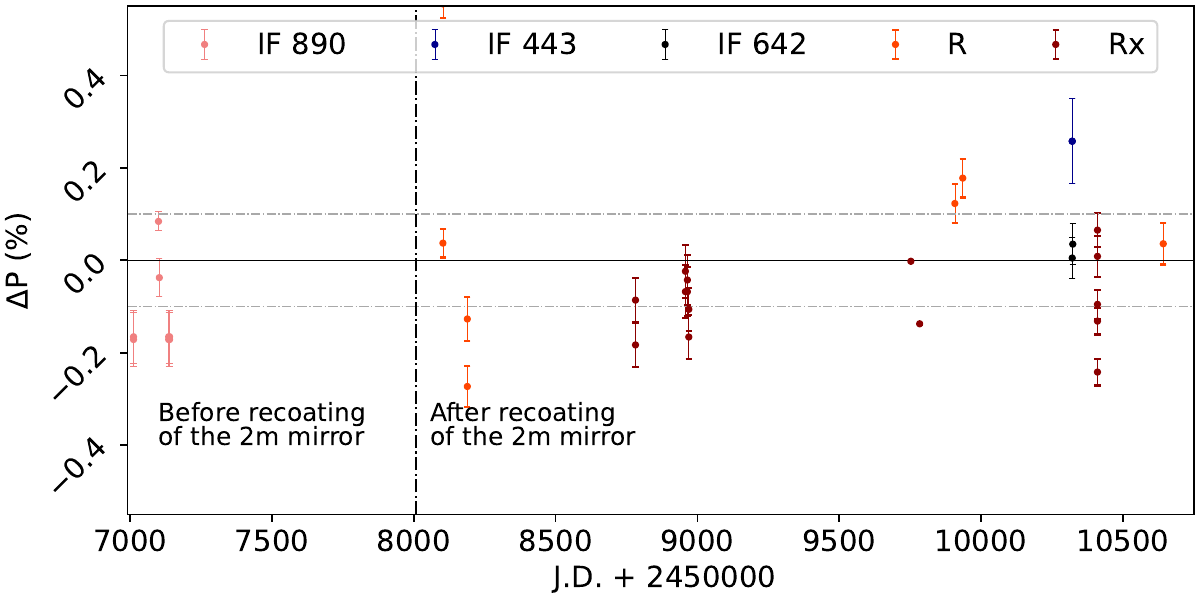}
   \end{center}
         \caption{Imaging polarimetry of unpolarised (left panel) and polarised (right panel) standard stars before and after the re-coating of the 2m mirror. The actual measurements are presented in Appendix~\ref{A:ImPol} in Tables \ref{TBL:ImPolZPol} and~\ref{TBL:ImPolHPol} therein.}
         \label{FIG:ImPol}
\end{figure*}

Imaging aperture polarimetry was performed as explained in \citet{2011JQSRT.112.2059B} by selecting the aperture at which the reduced Stokes parameters $P_Q = Q/I$ and $P_U = U/I$ converge to a well-defined value.
This method differs from the standard aperture photometry with apertures of increasing size in the way we chose the best aperture size. Here, we are looking at which aperture size the ratios between the fluxes in the parallel and the perpendicular beams (and, hence, the measured Stokes parameters) become constant. This aperture is usually smaller than the values where the individual fluxes become asymptotic. 
Technically, the best aperture can be selected based on the plots in Fig.~\ref{FIG:aper} and similar data.
The results of the imaging polarimetry measurements through the years of observations with the polarimetric mode of the FoReRo2 instrument are compared with their catalogue values and presented in Figure~\ref{FIG:ImPol} and Appendix~\ref{A:ImPol} within Tables~\ref{TBL:ImPolZPol} and~\ref{TBL:ImPolHPol} therein.

%
%
%
\section{Spectropolarimetric observations of high polarisation standards}
\label{spp.hpol}
In the spectropolarimetric mode of observation, we observed four unpolarised standard stars: HD 191195, HD 21447, HD 154892, and HD 212311. We also observed ten highly polarised standard stars: HD 7927, HD 25443, HD 161056, HD 19820, HD 204827, BD +25 727, HD 183143, BD +59 389, HD 154445, and HD 155197. The catalogue and observed values of these stars are presented in Table~\ref{tabl.list.sts} and their polarised spectra are presented in Figure~\ref{fig.serkowski}. The PA of the highly polarised standard stars is used as a reference. In Fig.~\ref{3D_pol}, we present 3D maps of these stars. The degree of polarisation is indicated by the length of the orange lines, while the orientation of each line represents the PA. The colour of each star corresponds to its distance.

Table~\ref{tabl.list.sts} includes the following information: the observed unpolarised and polarised standard stars, their coordinates, the number of observations for each star, the average value in the synthetic V band, the standard deviation, the degree of polarisation and PA from catalogues, and references. Two highly polarised standard stars, HD 183143 and HD 204827, exhibited variability during our observations.

\begin{table*}
\begin{threeparttable}
\caption{List of spectropolarimetric observed standard stars. }             
\begin{tabular}{l|c|c|c|c|c|cccc }     
\hline\hline       
Name  & R.A.\tnote{1}        & Dec.\tnote{1}     & \textnumero~of        & $\overline{P_V}$\tnote{2} &  $\sigma_{rms}(P_V)$\tnote{3} &   Band & P.D.  (\%) & PA (deg.)     &   References\\ 
      &  (J2000)    &  (J2000) & obs.     &    (\%)          &(\%)                  &        &  (catalogue)  &  (catalogue)       &                            \\
\hline     
\multicolumn{3}{l|}{\bf Unpolarised standard stars} & & & & & & & \\
HD 191195       & 20 06 13.9  & +53 09 56.5 & 10    &  0.03   & 0.02  &--&0.004 $\pm$ 0.001  &  &  (b) \\
HD 21447        & 03 30 00.2  & +55 27 06.5 & 16   &  0.05   & 0.02  &V &0.051  $\pm$ 0.020   &          &  (a) \\
HD 154892       & 17 07 41.3  & +15 12 37.6 & 20   &  0.08   & 0.04  &B &0.050  $\pm$ 0.030   &            &  (c) \\
HD 212311       & 22 21 58.6  & +56 31 52.7 & 10   &  0.11   & 0.03  &V &0.034  $\pm$ 0.021   &  &  (a) \\
\multicolumn{3}{l|}{\bf High polarisation standards} & & & & & & & \\
HD 7927     & 01 20 04.9  & +58 13 53.8 &  3    & 3.25   & 0.08   & V  &3.298  $\pm$ 0.025  &  ~~91.08 $\pm$ 0.20  & (a) \\
HD 25443    & 04 06 08.1  & +62 06 06.6 &  14   & 5.21   & 0.05   & V  &5.127  $\pm$ 0.061  & 134.23 $\pm$ 0.34  & (a) \\
HD 161056   & 17 43 47.0  & -07 04 46.6 &  27   & 4.05   & 0.04   & V  &4.030  $\pm$ 0.025  &  ~~66.93 $\pm$ 0.18  & (a) \\
HD 19820    & 03 14 05.3  & +59 33 48.5 &  11   & 4.82   & 0.03   & V  &4.787  $\pm$ 0.028  & 114.93 $\pm$ 0.17  & (a) \\
BD +25 727  & 04 44 24.9  & +25 31 42.4 &  2    & 6.31   &  --    & V  &6.280  $\pm$ 0.003  &  ~~32.20 $\pm$ 0.10  & (f) \\
BD +59 389  & 02 02 42.1  & +60 15 26.4 &  2    & 6.63   &  --    & V  &6.701  $\pm$ 0.015  &  ~~98.09 $\pm$ 0.07  & (a) \\
HD 154445   & 17 05 32.3  & -00 53 31.4 &  3    & 3.72   & 0.06   & V  &3.780  $\pm$ 0.062  &  ~~88.79 $\pm$ 0.47  & (a) \\
HD 155197   & 17 10 15.8  & -04 50 03.7 &  1    & 4.45   & --     & V  &4.320  $\pm$ 0.023  & 102.84 $\pm$ 0.15  & (a) \\
\multicolumn{3}{l|}{\bf Variable high polarisation standards} & & & & & & & \\
HD 183143   & 19 27 26.6  & +18 17 45.2 & 7   & 6.13   & \bf{0.15}      & V  & 6.100  $\pm$ 0.050  & 179.30   $\pm$ 0.20  & (d) \\
HD 204827   & 21 28 57.8  & +58 44 23.2 & 9   & 5.56   & \bf{0.20}      & V  & 5.322  $\pm$ 0.014  &  ~~58.73   $\pm$ 0.08  & (a) \\
            &             &             &     &        &            & V  & 5.607  $\pm$ 0.003  &  ~~58.33   $\pm$ 0.02  & (e) \\
\hline
\hline                                                                                                                  
\end{tabular}
References: (a) \text{\citet{1992AJ....104.1563S}}; (b) \text{\citet{2020A&A...635A..46P}}; (c) \text{\citet{1990AJ.....99.1243T}}; 
(d) \text{\citet{1982ApJ...262..732H}};\\(e) \text{\citet{2021AJ....161...20P}}; (f) Nordic Optical Telescope\tnote{4}
\begin{tablenotes}
\item[1] The coordinates were taken from the SIMBAD Astronomical Database.
\item[2] The average degree of polarisation in the synthetic V band.
\item[3] The standard deviation calculated as in Eq.~\ref{eq.sig}.
\item[4] Source: \url{https://www.not.iac.es/instruments/turpol/std/bd25727_note.html}.

\end{tablenotes}
\label{tabl.list.sts}   
\end{threeparttable}
\end{table*}

The average polarisation deviation between the catalogue values and the mean values measured for high-polarisation standard stars in the synthetic V band is ($\Delta{P_V}=|P_{V}(catalogue)-\overline{P_V}| = 0.06 \pm 0.04 \%$. This value is comparable to the standard deviation 
$\sigma_\mathrm{rms}(P_V)$ for the same set of stars, as listed in Table~\ref{tabl.list.sts}. Therefore, $\Delta P_V$ can be interpreted as a representative measure of the instrument’s polarimetric accuracy in the synthetic V band. The stars HD 183143 and HD 204827 were excluded from this calculation due to their known polarimetric variability.\\
Usually, the standard deviation is used to identify small fluctuations in magnitude and detect flickering (e.g., \citet{2023BlgAJ..38...83Z}). In this work, we applied the same approach to detect and confirm the variability of highly polarised standard stars. The standard deviation is calculated as 
\begin{equation}
    \sigma_{rms}(P_V) = \sqrt{ \frac{1}{N_{pts}-1} \sum\limits_{i}(P_V(i) - \overline P_V )^2 } ,
    \label{eq.sig}
\end{equation}
where $\overline P_V$ is the average degree of polarisation in the synthetic V band and $N_{pts}$ is the number of data points.
The $\sigma_{rms}$ calculated in this way includes the variability of the degree of polarisation of the star (if it exists).

In Table~\ref{tabl.list.sts}, the standard deviation is calculated for stars with more than two observations. The typical error of spectropolarimetric observations (presented in Figure~\ref{fig.serkowski}) is approximately 0.05–0.1\%. The standard deviations are comparable to this value. The standard deviation of unpolarised standard stars serves as an indicator of instrumental stability, while the standard deviation of highly polarised standard stars indicates the accuracy of the methodology.
Two of the observed highly polarised standard stars, HD 183143 and HD 204827, demonstrate a higher standard deviation: 0.15\% and 0.20\%, respectively. Both stars exhibit variable polarisation, which indicates the presence of an intrinsic component of polarisation.

Spectropolarimetric observations of HD 183143 with FoReRo2 span a period of two years, from 4 June 2022 to 24 October 2024. The star exhibits an $H_{\alpha}$ emission line and shows a variable degree of polarisation, indicative of an intrinsic polarisation component. The amplitude of variability in the synthetic V band is $P_{V}(max)-P_{V}(min)=0.38\%$. On each of the seven observing nights, another highly polarised standard star was observed alongside HD 183143. The PA of HD 183143 was corrected using the reference value for the corresponding night. Despite the variation in polarisation degree, the PA remains stable, with a mean value in the synthetic V band of $P.A._{V} = 177^\circ.6 \pm 0^\circ.6)$. Therefore, HD 183143 may still be used as a reference for PA calibration. It is worth noting that the PA we derived differs by approximately $2^\circ$ from the catalogue value, which might indicate long-term variability in the polarisation angle.

\citet{1986PASP...98..792D} concluded that HD 204827 possesses a component of intrinsic polarisation. Our observations confirm the variability of this star. HD 204827 was observed on two consecutive nights, 2 and 3 May 2025, with PA calibration performed using a second high-polarisation standard on each night (HD 161056). The applied PA corrections were $\Delta\theta = 5.43^\circ$ and $\Delta\theta = 5.61^\circ$ for 2 and 3 May, respectively. HD 204827 exhibits variability in both the degree of polarisation and the PA. In the synthetic V band, we obtained $P_V(\%) = 5.47 \pm 0.01$, $P.A._V = 61^\circ.8 \pm 0^\circ.1$ on 2 May 2025, and $P_V(\%) = 5.62 \pm 0.01$, $P.A._V = 58^\circ.0 \pm 0^\circ.1$ on 3 May 2025. Despite careful calibration, the derived PA of HD 204827 differed by nearly $4^\circ$ between the two nights, with no identifiable instrumental cause. This observed variability suggests that HD 204827 is not a reliable standard for polarimetric calibration, particularly with respect to the PA stability.

\section{FoReRo2 observational polarimetric examples}
\label{Sec.6}
As illustrative examples, we provide a statistical study of $K$ and \lmax based on the spectropolarimetric observations of polarised stars, spectropolarimetric observations of the recurrent nova RS Oph and the symbiotic star Z And, and the imaging polarimetry of the comet C/2019 Y4 (ATLAS).
\subsection{Serkowski's law: Relation between $K$ and \lmax}
\subsubsection{Fitting the polarisation spectra with Serkowski's law}
Serkowski's law is empirical and describes the wavelength dependence of linear polarisation \citep{1975ApJ...196..261S}, following 
\begin{equation}
P_{ISP}(\lambda ) = P_{max}  \; \exp \left(-K\ln ^{2}\frac{\lmax}{\lambda }\right),
\label{eq.Serkowski}
\end{equation}
where $P_{max}$ is the peak of the interstellar polarisation at wavelength \lmax. The K coefficient describes the shape of the curve and based on Serkowski's original work (Serkowski et al. 1975), it has a value of 1.15. The parameters of the Serkowski law fit are presented in Table~\ref{tab.Serkowski}. In Fig.~\ref{fig.serkowski}, the polarisation spectra of these high degree of polarisation standards are presented.

\begin{table*}[htb]
\centering
\caption{Parameters of the Serkowski law fit.}
\begin{tabular}{lccccc}
\hline
\bf High polarisation standards      & $P(\lambda)_{\text{max}}$ (\%) & $K$ & $\lambda_{\text{max}}$ ($\AA$) &$\overline{P_{V}}(\%)$\\ \hline        

\bf HD 7927        & $3.35 \pm 0.01$ & $1.09 \pm 0.01 $  & $5167 \pm  1.5$ & 3.25\\

\bf HD 25443       & $5.22 \pm 0.01$ & $1.05 \pm 0.01 $  & $5043 \pm  2.3$ & 5.21\\

\bf HD 161056      & $4.09 \pm 0.01$ & $1.37 \pm 0.01 $  & $5792 \pm  0.6$ & 4.05\\

\bf HD 19820       & $4.84 \pm 0.01$ & $1.07 \pm 0.01 $  & $5218 \pm  1.1$  & 4.82\\

\bf BD +25 727      & $6.36 \pm 0.01$ & $1.17 \pm 0.01 $  & $5689 \pm  1.3 $  & 6.31\\

\bf BD +59 389      & $6.66 \pm 0.01$ & $1.13 \pm 0.01 $  & $5613 \pm  0.8 $  & 6.63\\

\bf HD 154445      & $3.78 \pm 0.01$ & $1.28 \pm 0.01 $  & $5662 \pm  1.4 $  & 3.72\\

\bf HD 155197      & $4.48 \pm 0.01$ & $1.20 \pm 0.01 $  & $5664 \pm  4.4 $  & 4.45\\\hline
\bf Variable polarisation standards & $P(\lambda)_{\text{max}}$ (\%) & K & $\lambda_{\text{max}}$ ($\AA$) &${P_{V_{max}}}~and~{P_{V_{min}}}(\%)$\\
\hline
\bf HD 183143      &&&& \\
2023-08-13         & $6.32 \pm 0.01$ & $1.35 \pm 0.01 $  & $5547 \pm  0.6 $  & 6.3\\
2024-04-08         & $5.96 \pm 0.01$ & $1.41 \pm 0.01 $  & $5634 \pm  0.4 $  & 5.92\\
\hline
\bf HD 204827      &&&& \\ 
2019-07-02         & $5.82 \pm 0.01$ & $0.96 \pm 0.01 $  & $4502 \pm  3.7 $  & 5.67\\
2019-07-01         & $5.73 \pm 0.01$ & $0.60 \pm 0.01 $  & $3624 \pm  6.1 $  & 5.23\\
\hline
\label{tab.Serkowski}
\end{tabular} 
\\
\end{table*}
\subsubsection{A statistical study of $K$ and \lmax: Mean, standard deviation, and data distribution.}
\label{Sec.5}
Initially, K = 1.15 was assumed as a standard in the study by \citet{1973IAUS...52..145S}. An attempt to find a relation between the parameters of Serkowski's law, $K$, and \lmax\ can be found in the works of \citet{1982AJ.....87..695W,1992ApJ...386..562W}. In this section, we applied a statistical analysis of $K$ and \lmax\ to determine the mean, standard deviation, and data distribution. \citet{1992ApJ...386..562W} suggested the following relation, 
\begin{equation}
K = (0.01\pm0.05) + (1.66\pm0.09)\lmax,
\label{eq.klambda}
\end{equation}
where \lmax\ is expressed in $\mu$m.
\citet{2017MNRAS.464.4146C} compared the K and \lmax\ values presented in his paper with those reported by \citet{1992ApJ...386..562W}, revealing a systematic and statistically significant deviation, with the K values reported by \citet{1992ApJ...386..562W},  most likely being shifted due to incorrect effective wavelengths for the R and I passbands (see Section 4.2.1. in \citet{2017MNRAS.464.4146C}).

Since spectropolarimetric data do not suffer from the problem of properly characterising the photometric system \citep{2017MNRAS.464.4146C}, we decided to use only spectropolarimetric data published in the literature, together with data obtained with FR2, to perform a statistical analysis of the values in the $K-$\lmax\ plane.  We used data published in \citep{2017A&A...608A.146B, 2017MNRAS.464.4146C,2018A&A...615A..42C, 2022NewA...9701859N, 2023eas..conf.1422N}.
The coefficients $K$ and \lmax for non-variable standard stars are presented in Table~\ref{tab.Serkowski}. To calculate the mean and standard deviation, we used 146 stars, 31 of which were obtained with FR2. The mean of $\lambda_{max}$ is $\lambda_{max}=0.58 \mu m$ with a standard deviation of $\sigma_{\lmax}=0.07$\,$\mu$m. The mean of the K coefficient is 1.23 with a standard deviation of 0.32. It is noteworthy that the mean value of K is slightly higher than 1.15, the initial value published by \citet{1973IAUS...52..145S}.

\begin{figure*}[tbh]
\sidecaption
\includegraphics[width=12cm, angle=0]{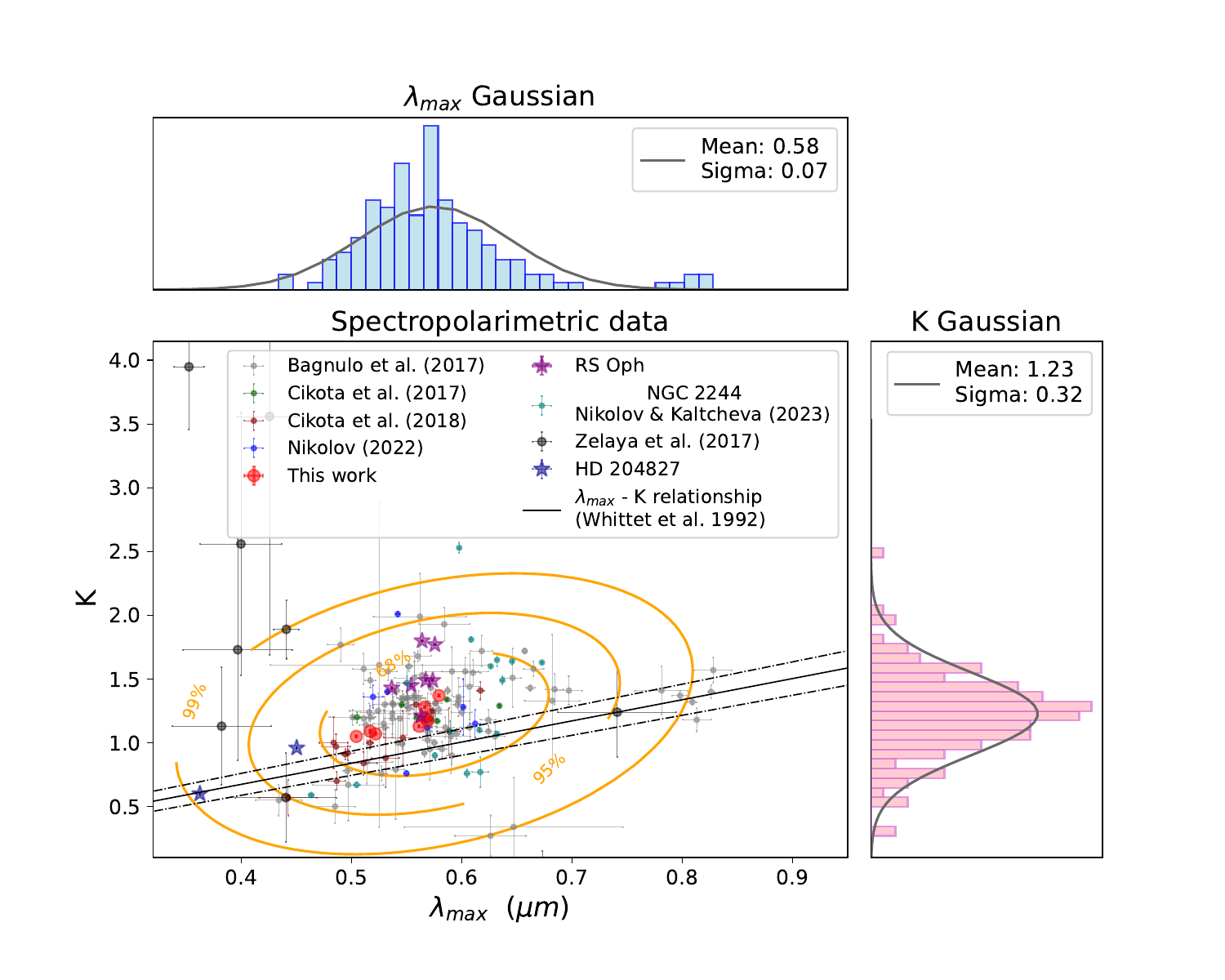}
\caption{$K$ and \lmax\ plane. The figure containing our eight unvariable standard stars compared to the sample from \citet{2017A&A...608A.146B, 2018A&A...615A..42C,2017MNRAS.464.4146C, 2022NewA...9701859N, 2017ApJ...836...88Z, 2023eas..conf.1422N}. The empirical $K-$\lmax\ relation determined by \citet{1992ApJ...386..562W} is presented with a black line and the dashed black lines trace the 1$\sigma$ uncertainty.}
\label{fig.klambda}
\end{figure*}

In Fig. \ref{fig.klambda}, we present $K-$\lmax\ values of Serkowski's fit of the spectropolarimetric data. We present the variable $K$ and \lmax\ of RS Oph (with purple) and HD 204827 (with dark blue).  The sample of SNe Ia \citep{2017ApJ...836...88Z} is shown in black, with details given in Appendix B of \citet{2018A&A...615A..42C}. We did not use these values to calculate the mean and standard deviation. In the $K-$\lmax~  plane, the spectropolarimetric standard stars (obtained in this work and \citet{2017MNRAS.464.4146C}) are located above the relation determined in \citet{1992ApJ...386..562W}.  Our observations of the standard stars (with red in Figure \ref{fig.klambda}) show the same systematic discrepancy. The orange ellipses present the areas with $1\sigma$; $2\sigma$, and $3\sigma$. Most of the data is located in $1\sigma$. Two of the values for the K coefficient of RS Oph (2 and 6 days after the most recent outburst) are located beyond $1\sigma$. The biggest deviation is observed in the sample of SN Ia \citep{2017ApJ...836...88Z}. 
%
%
\subsection{Variable $K$ and \lmax\ of the recurrent nova RS Oph.}
Spectropolarimetric observations of the recurrent nova RS Oph following its latest outburst indicate a significant intrinsic component of polarisation \citep{2023A&A...679A.150N}. In Table \ref{RSOph_table}, we present the fit of the spectropolarimetric observations of RS Oph using Serkowski's law. The peak value of the K coefficient is 1.80. From day 2 to day 7 after the 2021 outburst of RS Oph, the K coefficient decreases. It is noteworthy that the peak of the K coefficient within the first week of the most recent outburst of RS Oph corresponds to the newly formed dust during the event. This dust was destroyed within nine days of the 2021 outburst \citep{2023A&A...679A.150N}. In Table \ref{RSOph_table} we present the data with a correlation coefficient higher than 0.75. The variable K coefficient of RS Oph is presented in purple in Figure \ref{fig.klambda}.

The variation in RS Oph is dominated by changes in the Serkowski K coefficient, while 
\lmax\ remains essentially constant. This suggests that the intrinsic polarisation, due to nova-related dust, affects the shape of the polarisation curve without altering the peak wavelength. In contrast, some supernovae exhibit both high K values and unusually short \lmax\ (e.g., \citep{2017ApJ...836...88Z}, indicating either a different dust grain population or a more complex combination of intrinsic and interstellar polarisation components. Therefore, the behaviour of RS Oph cannot fully account for the polarimetric properties seen in supernovae, but it does provide a reference case for transient dust-induced intrinsic polarisation.

\begin{table*}
\centering
\caption{Variable $K$ and \lmax\ of the recurrent nova RS Oph.}
\label{RSOph_table}
\begin{tabular}{ l @{\hspace{0.5\tabcolsep}} | c @{\hspace{0.5\tabcolsep}} |c @{\hspace{0.5\tabcolsep}} |c @{\hspace{0.5\tabcolsep}} |c @{\hspace{0.5\tabcolsep}} |c @{\hspace{0.5\tabcolsep}} |c @{\hspace{0.5\tabcolsep}}}
\hline
\hline
 \multicolumn{1}{c@{\hspace{0.5\tabcolsep}}}{Date} & \multicolumn{1}{c@{\hspace{0.5\tabcolsep}}}{Days into the} & \multicolumn{1}{c@{\hspace{0.5\tabcolsep}}}{\lmax} & \multicolumn{1}{c@{\hspace{0.5\tabcolsep}}}{$P_{max}$} & \multicolumn{1}{c@{\hspace{0.5\tabcolsep}}}{K} & \multicolumn{1}{c@{\hspace{0.5\tabcolsep}}}{Correlation} & \multicolumn{1}{c@{\hspace{0.5\tabcolsep}}}{$P_{int}(V)$} \\
      &           \multicolumn{1}{c@{\hspace{0.5\tabcolsep}}}{2021 the outburst}       & \multicolumn{1}{c@{\hspace{0.5\tabcolsep}}}{$\mu$m} & \multicolumn{1}{c@{\hspace{0.5\tabcolsep}}}{$\%$}  &  &
       \multicolumn{1}{c@{\hspace{0.5\tabcolsep}}}{coefficient}
       & \multicolumn{1}{c@{\hspace{0.5\tabcolsep}}}{$\%$}
       \\
\hline
2019-07-02   &     -768   &       0.563 $\pm$  0.001    &          2.90 $\pm$    0.01 &       1.21 $\pm$    0.01  &  0.79  & $--$ \\
2021-08-10   &     2      &       0.564 $\pm$  0.000    &          3.09 $\pm$    0.01 &       1.80 $\pm$    0.01  &  0.96  & $1.20\pm0.04$ \\
2021-08-14   &     6      &       0.575 $\pm$  0.000    &          3.10 $\pm$    0.01 &       1.77 $\pm$    0.01  &  0.92  & $1.36\pm0.04$ \\
2021-08-15   &     7      &       0.567 $\pm$  0.000    &          3.08 $\pm$    0.01 &       1.49 $\pm$    0.01  &  0.92  & $1.21\pm0.04$\\
2022-06-04   &     300    &       0.554 $\pm$  0.001    &          2.93 $\pm$    0.01 &       1.45 $\pm$    0.01  &  0.78  & $0.38\pm0.05$ \\
2024-04-08   &     974    &       0.574 $\pm$  0.000    &          2.88 $\pm$    0.01 &       1.49 $\pm$    0.01  &  0.84  & $0.43 \pm 0.04$ \\
2024-05-05   &     1001   &       0.536 $\pm$  0.000    &          3.11 $\pm$    0.01 &       1.43 $\pm$    0.01  &  0.92  & $0.42 \pm 0.04$ \\
\hline
\end{tabular} 
\end{table*}
\subsection{Spectropolarimetric observation of symbiotic star Z And}
The spectropolarimetric observations of the two Raman OVI lines at $\lambda=6830\AA$ and $\lambda=7088\AA$ are powerful tools to diagnose the geometry of the symbiotic stars (\citet{1997A&A...321..791S}; \citet{1996A&A...310..235H}; \citet{1997A&A...327..219S}). These emission lines are due to Raman scattering of the OVI resonance doublet at $\lambda=1032\AA$ and $\lambda=1038\AA$ by neutral hydrogen (\citet{1989A&A...211L..31S}). Narrow-band Raman OVI imaging is typically used as a photometric tool for hunting 
symbiotic stars in the Local Universe (\citet{2019AJ....157..156A}). 

\begin{figure}[htb]
    \begin{center}
      \includegraphics[width=0.49\textwidth, angle=0]{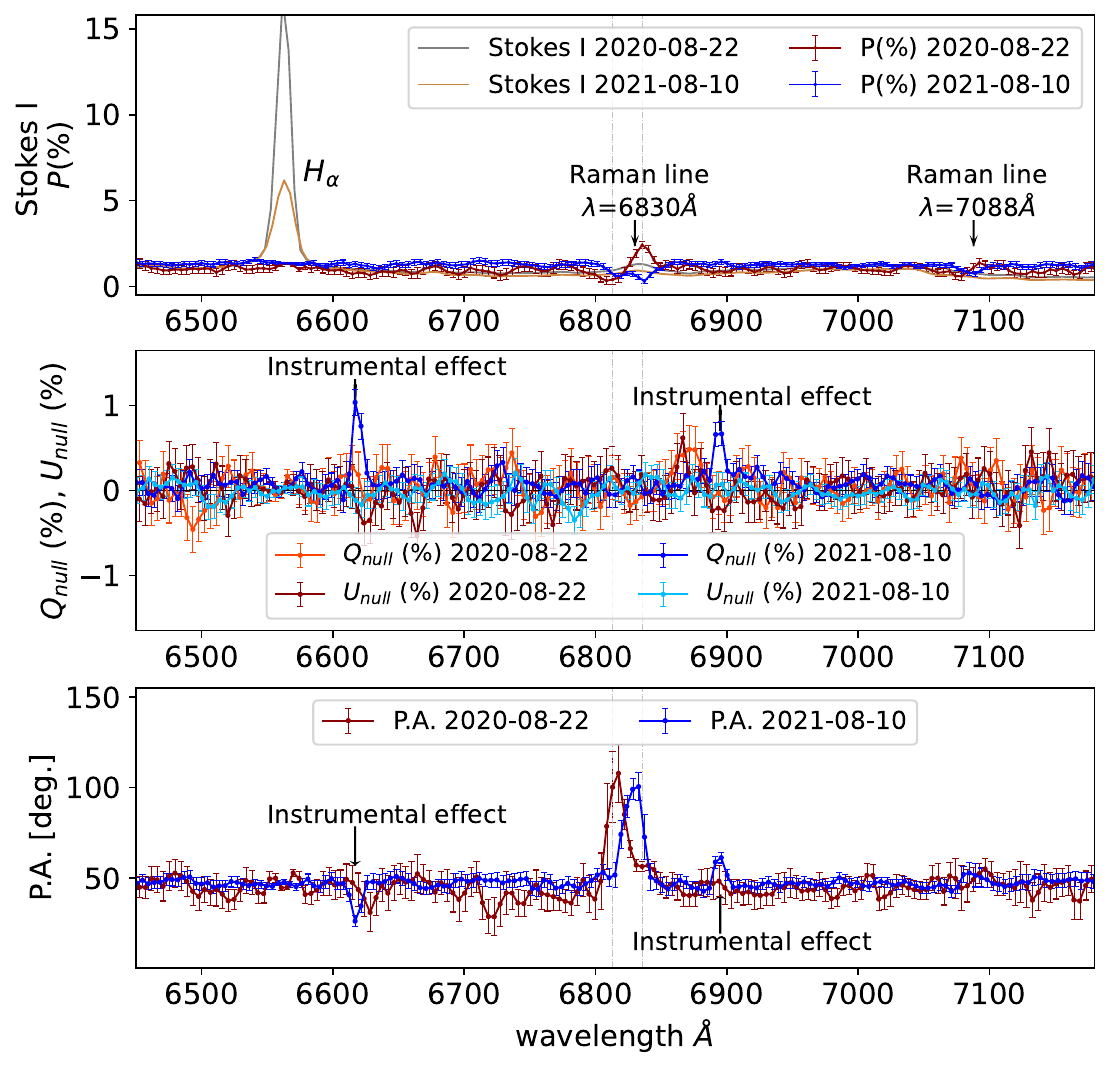}
   \end{center}
         \caption{Observed normalised intensity (Stokes I) and degree of polarisation (top panel). Null parameters (middle panel) and PA (bottom panel) of symbiotic star Z And.}
         \label{fig.ZAnd}
\end{figure}

Based on spectropolarimetric observations of the Raman lines, \citet{1997A&A...327..219S} determined the orbit inclination and the orbit orientation of Z And. We used this symbiotic star as a sample to explore the possibilities of FoReRo2 in the spectropolarimetric mode of observations. 
The symbiotic star Z And was observed on August 22, 2020 and August 10, 2021. In Fig. \ref{fig.ZAnd}, we show the normalised intensity (Stokes I) and observed degree of polarisation (top panel); null parameter (middle panel); and observed PA (bottom panel). The polarisation at $\lambda=6830\AA$ and $\lambda=7088\AA$ is amply visible. The null parameters are described in \citet{2009PASP..121..993B}. Both parameters were used as a quality check tool for data reduction. In Fig.~\ref{fig.ZAnd} (middle panel) both parameters: $Q_{null}$ and $U_{null}$ are close to zero at $\lambda=6830\AA$. That means the observed degree of polarisation is real.  

For the wavelength calibration, we can use Ne, HgCd, and He lamps. Our experience with the spectropolarimetric data reduction shows that wavelength calibration must be applied after splitting the spectra of o- and e-beam (both for object and arc lamp) into separate fit files. 
Otherwise, small wavelength shifts between o- and e-beam would lead to a dummy degree of polarisation in the spectral lines (for more information about small wavelength shifts and their impact on line polarisation, see \citet{2013A&A...559A.103B}).  
%
%
\subsection{Imaging polarimetry of comet C/2019 Y4 (ATLAS)}
Imaging polarimetry of comet C/2019 Y4 (ATLAS) in the wide-band filter Rx ($\lambda_0$=694.0\,nm, FWHM=79.0\,nm), which was chosen in such a way to avoid the molecular emission and only account for the reflected sunlight from the dust to pass through. It was carried out over 7 nights in April and May 2020 in a wide range of phase angles above 55$^{\circ}$ during its disintegration. To select the aperture for imaging polarimetry, we used the methodology described in Section~\ref{select.aperture}.

\begin{figure}[htb]
    \begin{center}
      \includegraphics[width=0.7\columnwidth, angle=90]{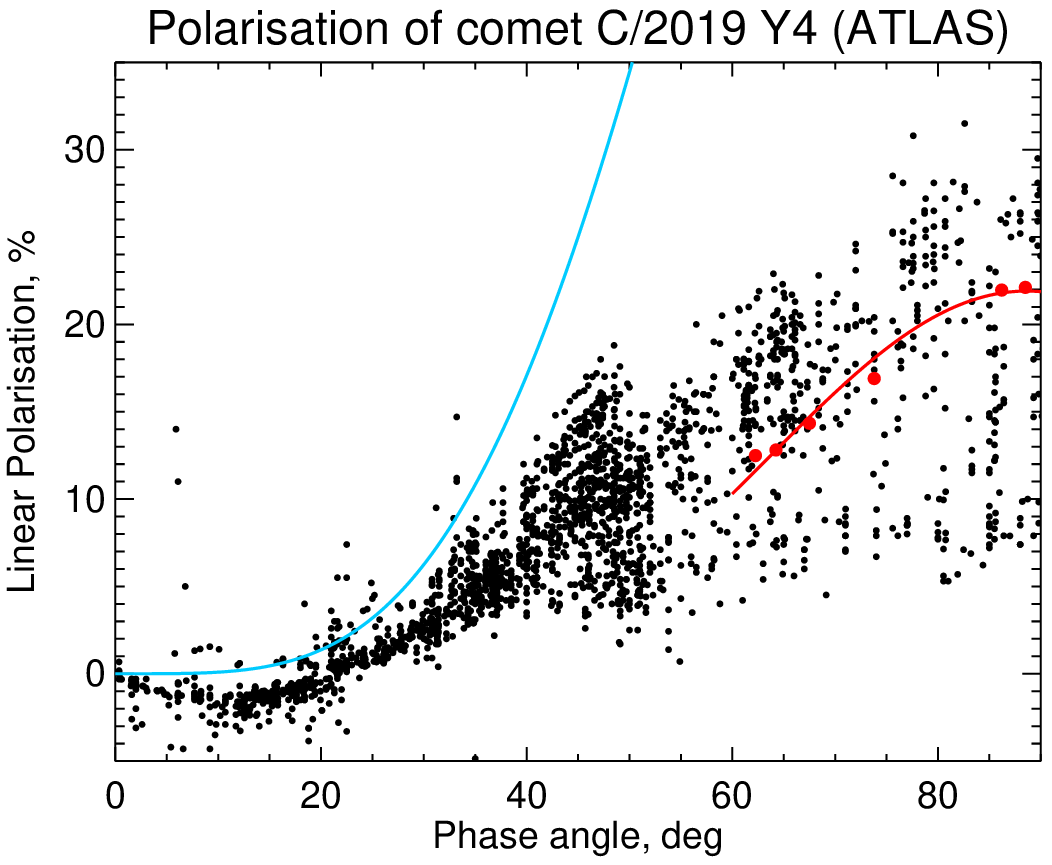}
   \end{center}
         \caption{Polarisation phase curve of the comet C/2019 Y4 (ATLAS). Red points are our observations and the red line is their fit with the function from Eq.~\ref{EQ:PolFit}. The blue line represents the Rayleigh scattering from a single particle with LP$_{\rm max}$=100\,\% at a phase angle of 90$^{\circ}$. Black points are measurements from the comets polarimetry database by \citet{2010PDSS.8124E....K}}
         \label{FIG:ATLAS}
\end{figure}

The measured degrees of linear polarisation for each night are presented in Table~\ref{TAB:ATLAS}. These results were fitted with the function from Eq.~\ref{EQ:PolFit} and presented together with measurements from the comets polarimetry database \citep{2010PDSS.8124E....K} in Fig.~\ref{FIG:ATLAS}. This function is expressed as
\begin{eqnarray}
\label{EQ:PolFit}
    {\rm LP}\left(\alpha\right)=b\left(\sin{\alpha}\right)^{c1}\left( \cos{\frac{\alpha}{2}} \right)^{c2}\sin{\left(\alpha-\alpha_0\right)}
.\end{eqnarray}
The obtained coefficients from the fit are as follows:
\[
\begin{array}{cclcl}
& & \alpha_0 & = & 20^{\circ}\,\,{\rm (fixed);}\\
& & b & = & 0.3292 \pm 0.0005;\\
& & c_1 & = & 4.0036 \pm 0.0324;\\
& & c_2 & = & 1\,\,{\rm (fixed).}
\end{array}
\]

Nevertheless, the disintegration of its nucleus, comet ATLAS, shows polarisation behaviour similar to any ordinary comet. Its phase-polarisation curve has a maximum linear polarisation LP$_{\rm max}$=21.9\,\% at phase angle of 88.6$^{\circ}$. 

The slope at the inversion angle is computed via Eq.~\ref{EQ:PolSlope} and gives h=$0.44$\,\%/deg, following 

\begin{eqnarray}
\label{EQ:PolSlope}
    {\rm h} = b\left(\sin{\alpha_0}\right)^{c1}\left( \cos{\frac{\alpha_0}{2}} \right)^{c2}
.\end{eqnarray}
\begin{table}
\centering
\caption{Imiging polarimetry of comet C/2019 Y4 (ATLAS)}
\label{TAB:ATLAS}
\begin{minipage}{1.0\columnwidth}\centering
\begin{tabular}{cccc}
\hline\hline 
Date\footnote{Date of observations, DD/MM/YYYY} & S-T-O\footnote{Phase angle (Sun-Target-Observer) in degrees} & LP\footnote{Linear polarisation in \%} & $\sigma_{\rm LP}$\footnote{1-$\sigma$ uncertainty of LP in \%} \\
\hline
18/04/2020 & 55.90 & 12.331 & 0.018 \\
25/04/2020 & 62.25 & 12.488 & 0.069 \\
27/04/2020 & 64.25 & 12.812 & 0.058 \\
30/04/2020 & 67.50 & 14.340 & 0.069 \\
05/05/2020 & 73.80 & 16.901 & 0.068 \\
12/05/2020 & 86.20 & 21.965 & 0.041 \\
13/05/2020 & 88.50 & 22.119 & 0.060 \\
\end{tabular} 
\end{minipage}
\end{table}
\section{Summary and conclusions}
In this study, we present an overview of the 2-Channel Focal Reducer Rozhen (FoReRo2), with an emphasis on its polarimetric and spectropolarimetric observing modes. To assess the instrument’s performance and stability, we analysed spectropolarimetric observations of four unpolarised standard stars over 62 epochs, along with ten high-polarisation standard stars over 79 epochs. The dataset spans a decade of polarimetric and spectropolarimetric observations with FoReRo2.
\begin{itemize}
     \item The typical polarimetric accuracy of FoReRo2, estimated from high-polarisation standard stars, is better than 0.1\% in the synthetic V band and in the red domain from imaging polarimetry.
     \item The standard deviation on long-term observations of high polarisation standards can be used as a tool to detect the variability of the degree of polarisation. We used this methodology to demonstrate the variability of two high-polarisation standard stars: HD 183143 and HD 204827.
     \item Among the ten high-polarisation standard stars analysed in this work, HD 204827 exhibited significant short-term variability in its PA. Based on our observations, we conclude that HD 204827 is unsuitable for use as a reliable polarimetric standard.
     \item Based on spectropolarimetric observations, we made a statistical analysis of the Serkowski law parameters (\lmax\ and $K$). We obtained the mean of \lmax\ as 0.58\,$\mu m$ with a standard deviation of $\sigma_{\lambda_{max}}=0.07 \mu m$. The mean of the $K$ coefficient is 1.23 with a standard deviation of 0.32. 
     \item The selection of the aperture for imaging polarimetry was done by choosing the aperture where the reduced Stokes parameters ($P_Q = Q/I$ and $P_U = U/I$) converge to a well-defined value. 
     \item  Only slitless spectropolarimetry gives reasonable results, while normal slit spectropolarimetry seems to be affected by strong instrumental polarisation. This is likely due to a slight misalignment of the 2-meter mirror relative to the optical axis, which appeared when it was mounted after its last re-coating.
     \item The 2021 outburst of RS Oph demonstrates that intrinsic polarisation caused by nova-related dust can modify the Serkowski K parameter, with no significant change in 
     \lmax. This provides a reference case for dust-induced polarimetric variability in transient events. 
     \item Symbiotic star Z And demonstrates variable polarisation in the Raman OVI line at $\lambda=6830\AA$.
     \item Disintegrated comet ATLAS shows maximum linear polarisation LP$_{\rm max}$=21.9\,\% at a phase angle of 88.6$^{\circ}$, similarly to any ordinary comet. 
\end{itemize} 
To our knowledge, FoReRo2 is the only operational low-resolution optical spectropolarimeter mounted on a 2-meter-class telescope in Southeastern Europe. This work presents the first comprehensive calibration of the instrument and demonstrates its long-term stability and scientific applicability. The results provide a reliable reference baseline for future polarimetric and spectropolarimetric observations with FoReRo2. The instrument is well-suited for transient phenomena, as well as for regular studies of solar system bodies, the interstellar medium, and various stellar sources.
\begin{acknowledgements}
Y.N. acknowledges partial support by grants: K$\Pi$-06-H58/3 "Space weather and habitability of exoplanets: the role of coronal mass ejections and (super)flares" and K$\Pi$-06-M88/1 "Interstellar medium - 3D distribution and physical characteristics", by the Bulgarian National Science Fund. 

G.B., Y.N. and T.B. acknowledge partial support by grants: K$\Pi$-06-H88/5 “Physical properties and chemical composition of asteroids and comets - a key to increasing our knowledge of the Solar System origin and evolution.” by the Bulgarian National Science Fund.

Yanko Nikolov, thanks for the hospitality of Armagh Observatory and Planetarium, and for the opportunity to work there for a period of several weeks. The research that led to these results was partially carried out with the help of infrastructure purchased under the National Roadmap for Research Infrastructure, financially coordinated by the Ministry of Education and Science of Republic of Bulgaria.

Astrometry.net was used to derive the scale of the system with the new CCD cameras.
 
This research has made use of NASA’s Astrophysics Data System Bibliographic Services.
\end{acknowledgements}
\bibliographystyle{aa} 
\bibliography{aa58243-25}

\begin{appendix}
\label{sec:Standard stars}
\section{Imaging polarimetry of standard stars}\label{A:ImPol}
This appendix presents imaging polarimetry results for both unpolarised and polarised standard stars. Table~\ref{TBL:ImPolZPol} summarises the measurements for unpolarised standard stars, while Table~\ref{TBL:ImPolHPol} lists the results for polarised standard stars. The data illustrate the instrument’s performance in imaging polarimetric mode across various bands. Some observations were conducted with narrowband filters.
\begin{table}[H]
\centering
\caption{\mbox{Results from imaging polarimetry of unpolarised standard stars.}}
\resizebox{\textwidth}{!}{%
\begin{tabular}{lllccccrcl}  
\hline
\hline
Star      &  Date        &  Filter  &  P,       &  $\sigma$P,      &  P$_{\rm cat}$, &  $\sigma$P$_{\rm cat}$, & \multicolumn{1}{c}{(P-P$_{\rm cat}$),}  &  $\sigma$(P-P$_{\rm cat}$),  &  Notes \\
 & & & \% & \% & \% & \% & \multicolumn{1}{c}{\%} & \% & \\
\hline
HD42807   &  2014-12-20  &  V       &  0.07  &  0.01  &  0.00  &  0.00  &   0.07  &  0.01  &  UKIRT; 000$\dots$067 \\
G191B2B   &  2017-12-15  &  R       &  0.13  &  0.05  &  0.09  &  0.05  &   0.04  &  0.07  &  NOT; 000$\dots$067 \\
G191B2B   &  2017-12-15  &  R       &  0.20  &  0.09  &  0.09  &  0.05  &   0.11  &  0.10  &  NOT; 090$\dots$157 \\
HD154892  &  2018-03-11  &  R       &  0.07  &  0.00  &  0.05  &  0.03  &   0.02  &  0.03  &  NOT; 000$\dots$067 \\
HD154892  &  2018-03-11  &  R       &  0.12  &  0.00  &  0.05  &  0.03  &   0.07  &  0.03  &  NOT; 090$\dots$157 \\
--        &  2018-10-02  &  R       &  0.04  &  0.03  &  0.00  &  0.00  &   0.04  &  0.03  &  000$\dots$067 \\
G191B2B   &  2019-09-21  &  Rx      &  0.13  &  0.05  &  0.09  &  0.05  &   0.04  &  0.07  &  NOT; 000$\dots$067 \\
G191B2B   &  2019-09-21  &  Rx      &  0.09  &  0.05  &  0.09  &  0.05  &  -0.01  &  0.07  &  NOT; 090$\dots$157 \\
G191B2B   &  2019-10-21  &  Rx      &  0.11  &  0.05  &  0.09  &  0.05  &   0.02  &  0.07  &  NOT; 000$\dots$067 \\
G191B2B   &  2019-10-21  &  Rx      &  0.20  &  0.05  &  0.09  &  0.05  &   0.11  &  0.07  &  NOT; 090$\dots$157 \\
G191B2B   &  2019-10-22  &  Rx      &  0.03  &  0.05  &  0.09  &  0.05  &  -0.06  &  0.07  &  NOT; 000$\dots$067 \\
G191B2B   &  2019-10-22  &  Rx      &  0.09  &  0.05  &  0.09  &  0.05  &  -0.01  &  0.07  &  NOT; 090$\dots$157 \\
G191B2B   &  2019-10-24  &  Rx      &  0.11  &  0.06  &  0.09  &  0.05  &   0.02  &  0.08  &  NOT; 000$\dots$067 \\
G191B2B   &  2019-10-24  &  Rx      &  0.06  &  0.06  &  0.09  &  0.05  &  -0.03  &  0.08  &  NOT; 090$\dots$157 \\
G191B2B   &  2019-10-25  &  Rx      &  0.09  &  0.07  &  0.09  &  0.05  &  -0.00  &  0.09  &  NOT; 000$\dots$067 \\
G191B2B   &  2019-10-25  &  Rx      &  0.06  &  0.08  &  0.09  &  0.05  &  -0.03  &  0.09  &  NOT; 090$\dots$157 \\
G191B2B   &  2020-04-18  &  Rx      &  0.10  &  0.06  &  0.09  &  0.05  &   0.01  &  0.07  &  NOT; 000$\dots$067 \\
G191B2B   &  2020-04-18  &  Rx      &  0.12  &  0.06  &  0.09  &  0.05  &   0.03  &  0.07  &  NOT; 090$\dots$157 \\
G191B2B   &  2020-04-25  &  Rx      &  0.05  &  0.05  &  0.09  &  0.05  &  -0.04  &  0.07  &  NOT; 000$\dots$067 \\
G191B2B   &  2020-04-25  &  Rx      &  0.04  &  0.05  &  0.09  &  0.05  &  -0.06  &  0.07  &  NOT; 090$\dots$157 \\
HD154892  &  2020-04-30  &  Rx      &  0.08  &  0.04  &  0.05  &  0.03  &   0.03  &  0.05  &  NOT; 000$\dots$067 \\
HD154892  &  2020-04-30  &  Rx      &  0.05  &  0.04  &  0.05  &  0.03  &   0.00  &  0.05  &  NOT; 090$\dots$157 \\
HD154892  &  2020-05-01  &  Rx      &  0.06  &  0.03  &  0.05  &  0.03  &   0.01  &  0.04  &  NOT; 000$\dots$157 \\
--        &  2022-06-04  &  R       &  0.06  &  0.04  &  0.00  &  0.00  &   0.06  &  0.04  &  000$\dots$067 \\
--        &  2022-06-04  &  R       &  0.01  &  0.03  &  0.00  &  0.00  &   0.01  &  0.03  &  090$\dots$157 \\
HD154892  &  2022-06-22  &  Rx      &  0.18  &  0.03  &  0.05  &  0.03  &   0.13  &  0.04  &  NOT; 000$\dots$157 \\
HD154892  &  2022-07-23  &  Rx      &  0.20  &  0.04  &  0.05  &  0.03  &   0.15  &  0.05  &  NOT; 000$\dots$157 \\
G191B2B   &  2022-11-24  &  R       &  0.06  &  0.06  &  0.09  &  0.05  &  -0.03  &  0.07  &  NOT; 000$\dots$157 \\
G191B2B   &  2022-12-22  &  R       &  0.08  &  0.10  &  0.09  &  0.05  &  -0.01  &  0.11  &  NOT; 000$\dots$157 \\
HD65629   &  2024-01-12  &  IF443   &  0.31  &  0.04  &  0.01  &  0.00  &   0.31  &  0.04  &  NOT; 000$\dots$157 \\
HD65629   &  2024-01-12  &  IF684   &  0.48  &  0.03  &  0.01  &  0.00  &   0.48  &  0.03  &  NOT; 000$\dots$157 \\
G191B2B   &  2024-11-28  &  R       &  0.05  &  0.03  &  0.09  &  0.05  &  -0.04  &  0.06  &  NOT; 000$\dots$157 \\
G191B2B   &  2024-12-29  &  Rx      &  0.12  &  0.04  &  0.09  &  0.05  &   0.03  &  0.06  &  NOT; 000$\dots$157 \\
\hline
\end{tabular}%
}
\label{TBL:ImPolZPol}
\end{table}
\begin{table*}
\caption{Results from imaging polarimetry of polarised standard stars.}    \begin{tabular}{lllccccrcl}  
\hline
\hline
Star      &  Date        &  Filter  &  P,       &  $\sigma$P,      &  P$_{\rm cat}$, &  $\sigma$P$_{\rm cat}$, &  \multicolumn{1}{c}{(P-P$_{\rm cat}$),}  &  $\sigma$(P-P$_{\rm cat}$),  &  Notes \\
 & & & \% & \% & \% & \% & \multicolumn{1}{c}{\%} & \% & \\
\hline
HD183143    &  2014-12-20  &  IF890  &  4.56  &  0.06  &  4.72  &   --   &  -0.17  &  0.06  &  UKIRT; 000$\dots$067 \\
HD183143    &  2014-12-20  &  IF890  &  4.55  &  0.06  &  4.72  &   --   &  -0.17  &  0.06  &  UKIRT; 090$\dots$157 \\
HD43384     &  2015-03-18  &  IF890  &  2.30  &  0.02  &  2.22  &   --   &   0.08  &  0.02  &  UKIRT; 000$\dots$067 \\
HD183143    &  2015-03-21  &  IF890  &  4.68  &  0.04  &  4.72  &   --   &  -0.04  &  0.04  &  UKIRT; 000$\dots$067 \\
HD183143    &  2015-04-23  &  IF890  &  4.56  &  0.06  &  4.72  &   --   &  -0.17  &  0.06  &  UKIRT; 000$\dots$067 \\
HD183143    &  2015-04-23  &  IF890  &  4.55  &  0.06  &  4.72  &   --   &  -0.17  &  0.06  &  UKIRT; 090$\dots$157 \\
HD183143    &  2015-04-26  &  IF890  &  4.56  &  0.06  &  4.72  &   --   &  -0.17  &  0.06  &  UKIRT; 000$\dots$067 \\
HD183143    &  2015-04-26  &  IF890  &  4.55  &  0.06  &  4.72  &   --   &  -0.17  &  0.06  &  UKIRT; 090$\dots$157 \\
BDp59389    &  2017-12-15  &  R      &  6.47  &  0.02  &  6.43  &  0.02  &   0.04  &  0.03  &  NOT; 000$\dots$067 \\
BDp59389    &  2017-12-15  &  R      &  6.99  &  0.02  &  6.43  &  0.02  &   0.56  &  0.03  &  NOT; 090$\dots$157 \\
HD161056    &  2018-03-11  &  R      &  3.89  &  0.04  &  4.01  &  0.03  &  -0.13  &  0.05  &  NOT; 000$\dots$067 \\
HD161056    &  2018-03-11  &  R      &  3.74  &  0.03  &  4.01  &  0.03  &  -0.27  &  0.04  &  NOT; 090$\dots$157 \\
BDp59389    &  2019-10-25  &  Rx     &  6.34  &  0.04  &  6.43  &  0.02  &  -0.09  &  0.05  &  NOT; 000$\dots$067 \\
BDp59389    &  2019-10-25  &  Rx     &  6.25  &  0.04  &  6.43  &  0.02  &  -0.18  &  0.05  &  NOT; 090$\dots$157 \\
HD25443     &  2020-04-18  &  Rx     &  4.71  &  0.04  &  4.73  &  0.05  &  -0.02  &  0.06  &  NOT; 000$\dots$067 \\
HD25443     &  2020-04-18  &  Rx     &  4.67  &  0.04  &  4.73  &  0.05  &  -0.07  &  0.06  &  NOT; 090$\dots$157 \\
HD25443     &  2020-04-25  &  Rx     &  4.67  &  0.03  &  4.73  &  0.05  &  -0.07  &  0.05  &  NOT; 000$\dots$067 \\
HD25443     &  2020-04-25  &  Rx     &  4.69  &  0.03  &  4.73  &  0.05  &  -0.04  &  0.05  &  NOT; 090$\dots$157 \\
HD161056    &  2020-04-30  &  Rx     &  3.91  &  0.03  &  4.01  &  0.03  &  -0.11  &  0.05  &  NOT; 000$\dots$067 \\
HD161056    &  2020-04-30  &  Rx     &  3.85  &  0.03  &  4.01  &  0.03  &  -0.17  &  0.05  &  NOT; 090$\dots$157 \\
HD161056    &  2022-06-22  &  Rx     &  4.01  &  0.00  &  4.01  &  0.00  &  -0.00  &  0.00  &  NOT; 000$\dots$157 \\
HD161056    &  2022-07-23  &  Rx     &  3.87  &  0.00  &  4.01  &  0.00  &  -0.14  &  0.00  &  NOT; 000$\dots$157 \\
HD251204    &  2022-11-24  &  R      &  4.84  &  0.00  &  4.72  &  0.04  &   0.12  &  0.04  &  NOT; 000$\dots$157 \\
HD251204    &  2022-12-22  &  R      &  4.90  &  0.00  &  4.72  &  0.04  &   0.18  &  0.04  &  NOT; 000$\dots$157 \\
HD25443     &  2024-01-12  &  IF443  &  5.49  &  0.00  &  5.23  &  0.09  &   0.26  &  0.09  &  NOT; 000$\dots$157 \\
HD25443     &  2024-01-12  &  IF642  &  4.74  &  0.00  &  4.73  &  0.05  &   0.01  &  0.05  &  NOT; 000$\dots$157 \\
HD25443     &  2024-01-14  &  IF642  &  4.77  &  0.00  &  4.73  &  0.05  &   0.04  &  0.05  &  NOT; 000$\dots$157 \\
HD204827    &  2024-04-10  &  Rx     &  4.76  &  0.00  &  4.89  &  0.03  &  -0.13  &  0.03  &  NOT; 000$\dots$157 \\
HD161056    &  2024-04-10  &  Rx     &  3.92  &  0.00  &  4.01  &  0.03  &  -0.10  &  0.03  &  NOT; 000$\dots$157 \\
Hiltner960  &  2024-04-10  &  Rx     &  4.97  &  0.00  &  5.21  &  0.03  &  -0.24  &  0.03  &  NOT; 000$\dots$157 \\
VICyg12     &  2024-04-10  &  Rx     &  7.96  &  0.00  &  7.89  &  0.04  &   0.07  &  0.04  &  NOT; 000$\dots$157 \\
HD283812    &  2024-04-10  &  Rx     &  4.74  &  0.00  &  4.73  &  0.05  &   0.01  &  0.05  &  NOT; 000$\dots$157 \\
HD283812    &  2024-11-28  &  R      &  4.77  &  0.00  &  4.73  &  0.05  &   0.04  &  0.05  &  NOT; 000$\dots$157 \\
\hline
\end{tabular}
\label{TBL:ImPolHPol}
\end{table*}
\FloatBarrier
\section{Polarised spectra of high-polarisation standards.}
\begin{figure*}[htb]
    \begin{center}
      \includegraphics[width=0.9\textwidth, angle=0]{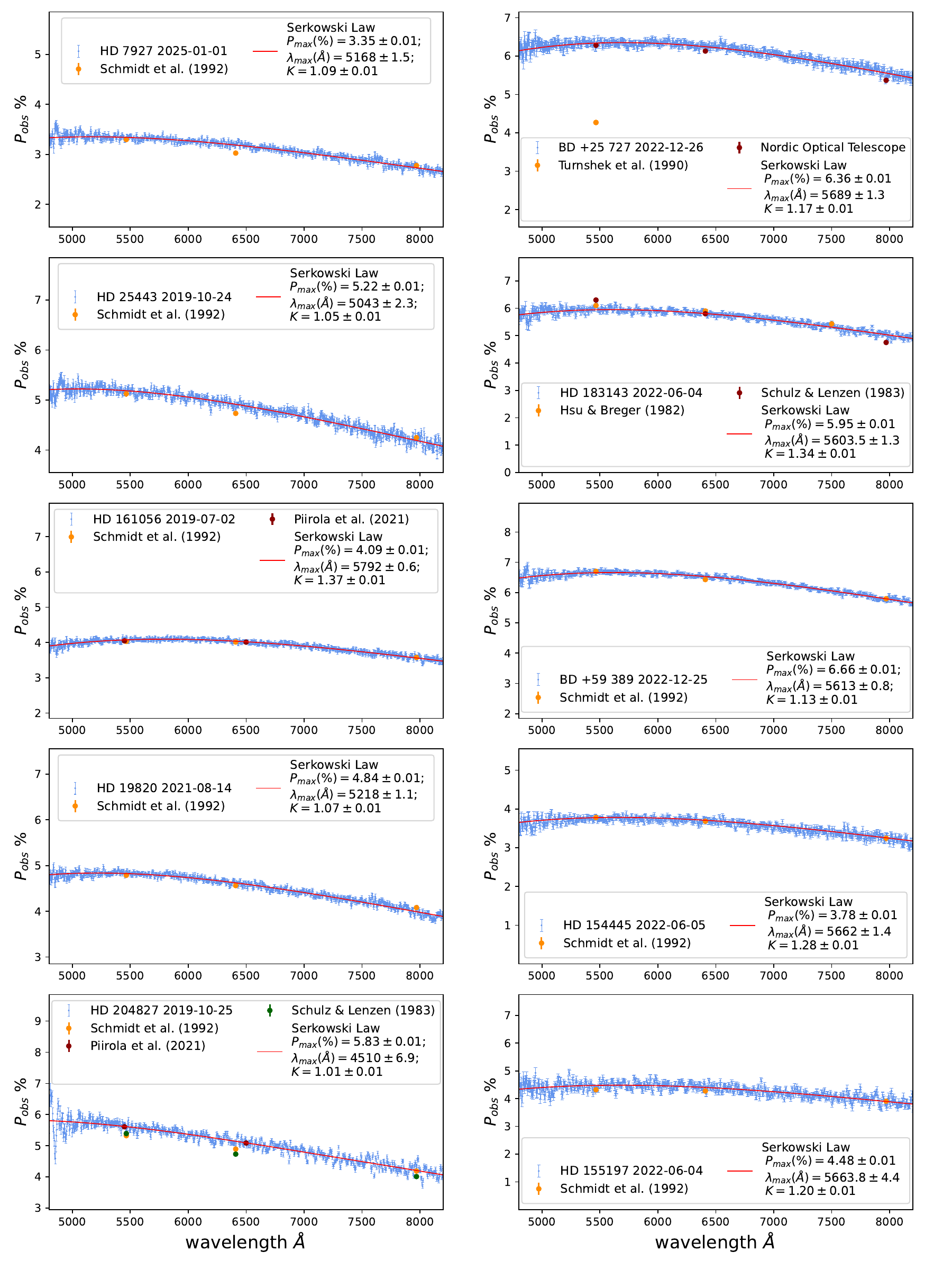}
   \end{center}
         \caption{Polarised spectra of high-polarisation standards HD7927, HD25443, HD161056, HD 19820, HD 204827, BD +25 727, HD 183143, BD +59 389, HD 154445, and HD 155197.}
         \label{fig.serkowski}
\end{figure*}

In this appendix, we present the polarised spectra of the high-polarisation standard stars: HD 7927, HD 25443, HD 161056, HD 19820, HD 204827, BD +25 727, HD 183143, BD +59 389, HD 154445, and HD 155197 (see Fig.~\ref{fig.serkowski}). The orange, red, and green dots represent the catalogue values taken from \citet{1992AJ....104.1563S, 2021AJ....161...20P, 1983A&A...121..158S, 1990AJ.....99.1243T, 1982ApJ...262..732H}, and the data presented on the Nordic Optical Telescope's website\footnotemark. The red curve represents Serkowski's law, and the coefficients of Serkowski's law are presented in Table~\ref{tab.Serkowski}.
\footnotetext{Source: \url{https://www.not.iac.es/instruments/turpol/std/bd25727_note.html}}
%
%
\section{3D polarisation maps of high-polarisation standards.}
\label{3D_hpol}
\begin{figure*}[!htb]
    \centering    
    \subfloat{\includegraphics[width=0.33\textwidth, angle=0]{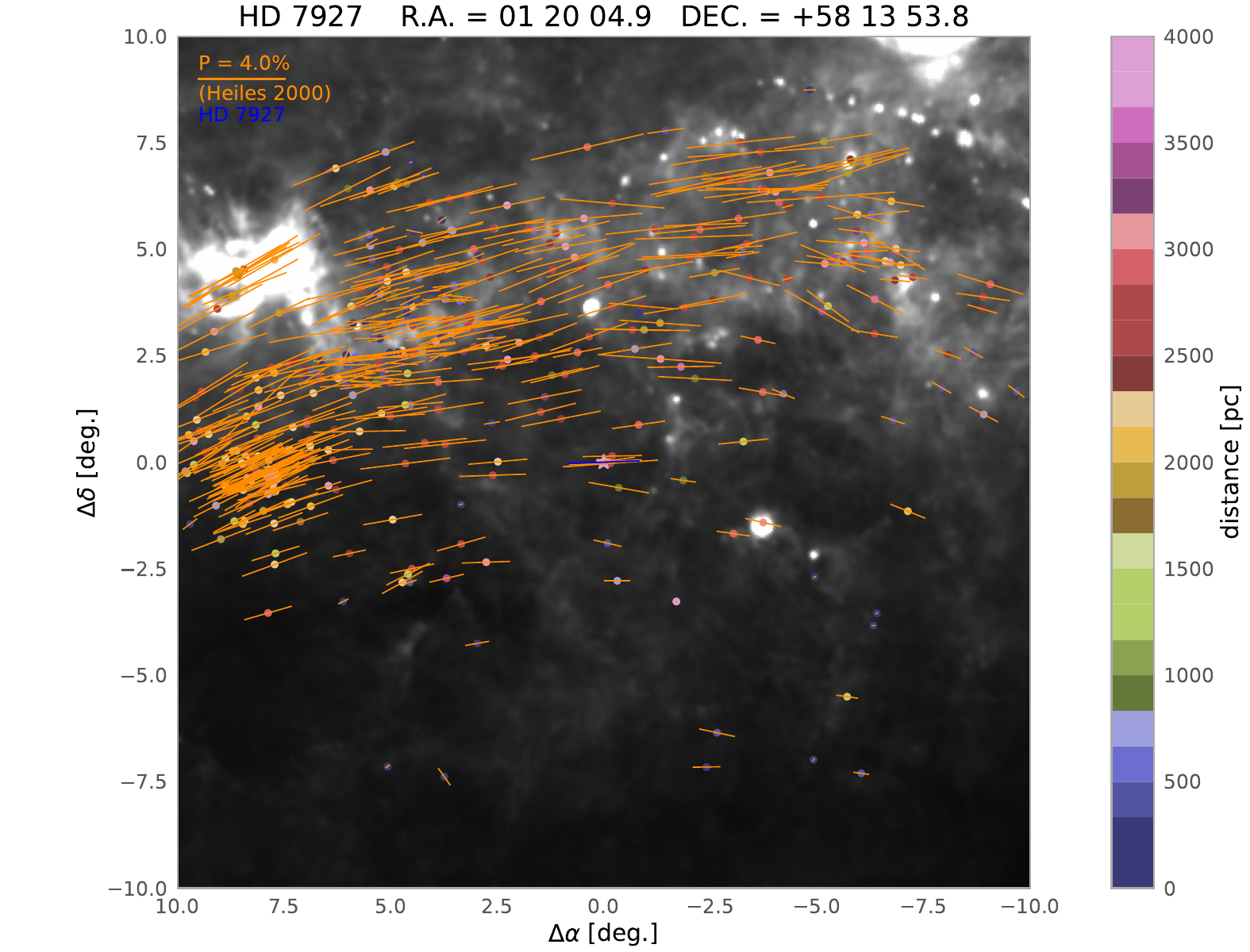}}\
    \subfloat{\includegraphics[width=0.33\textwidth, angle=0]{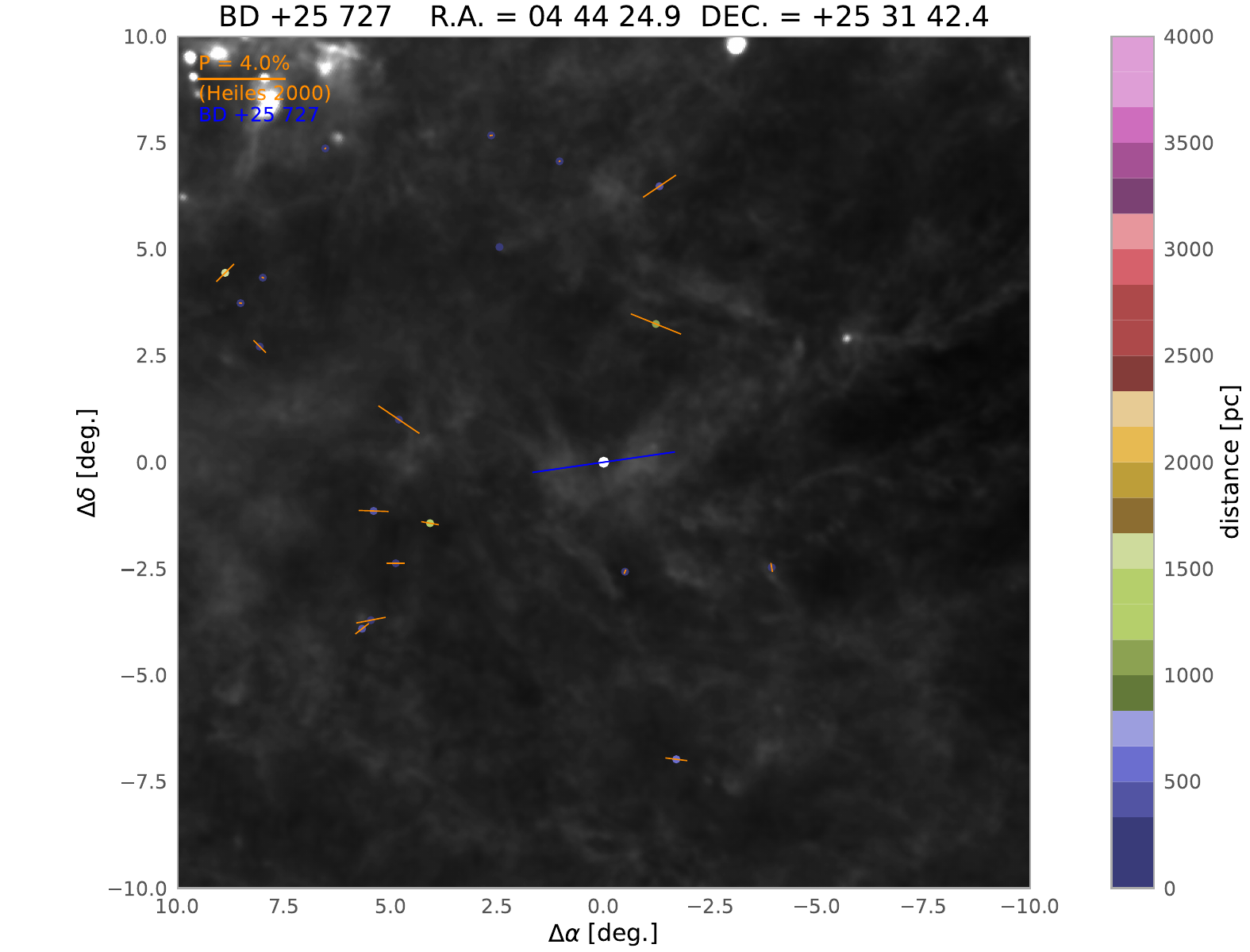}}\
    \subfloat{\includegraphics[width=0.33\textwidth, angle=0]{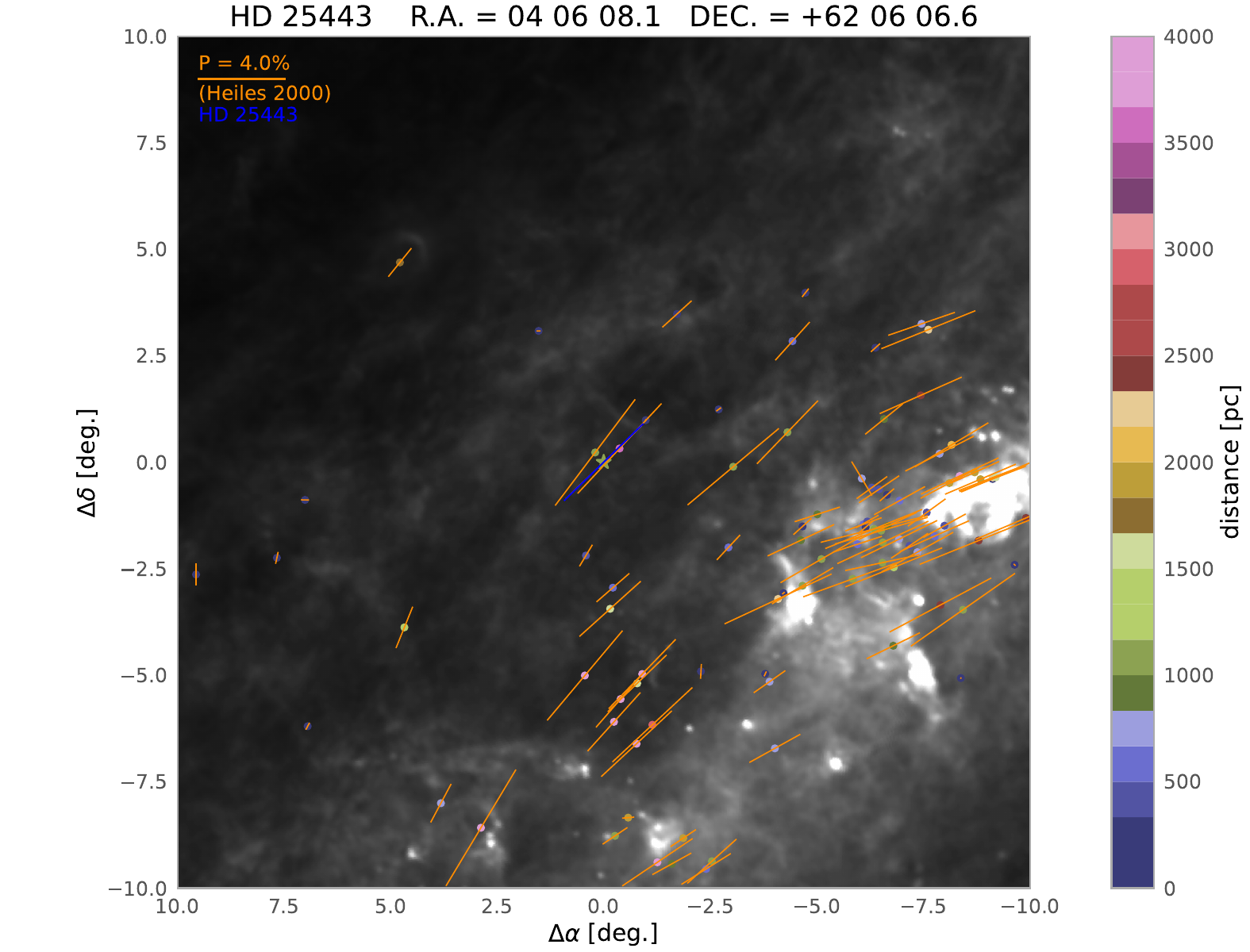}}\\
    \subfloat{\includegraphics[width=0.33\textwidth, angle=0]{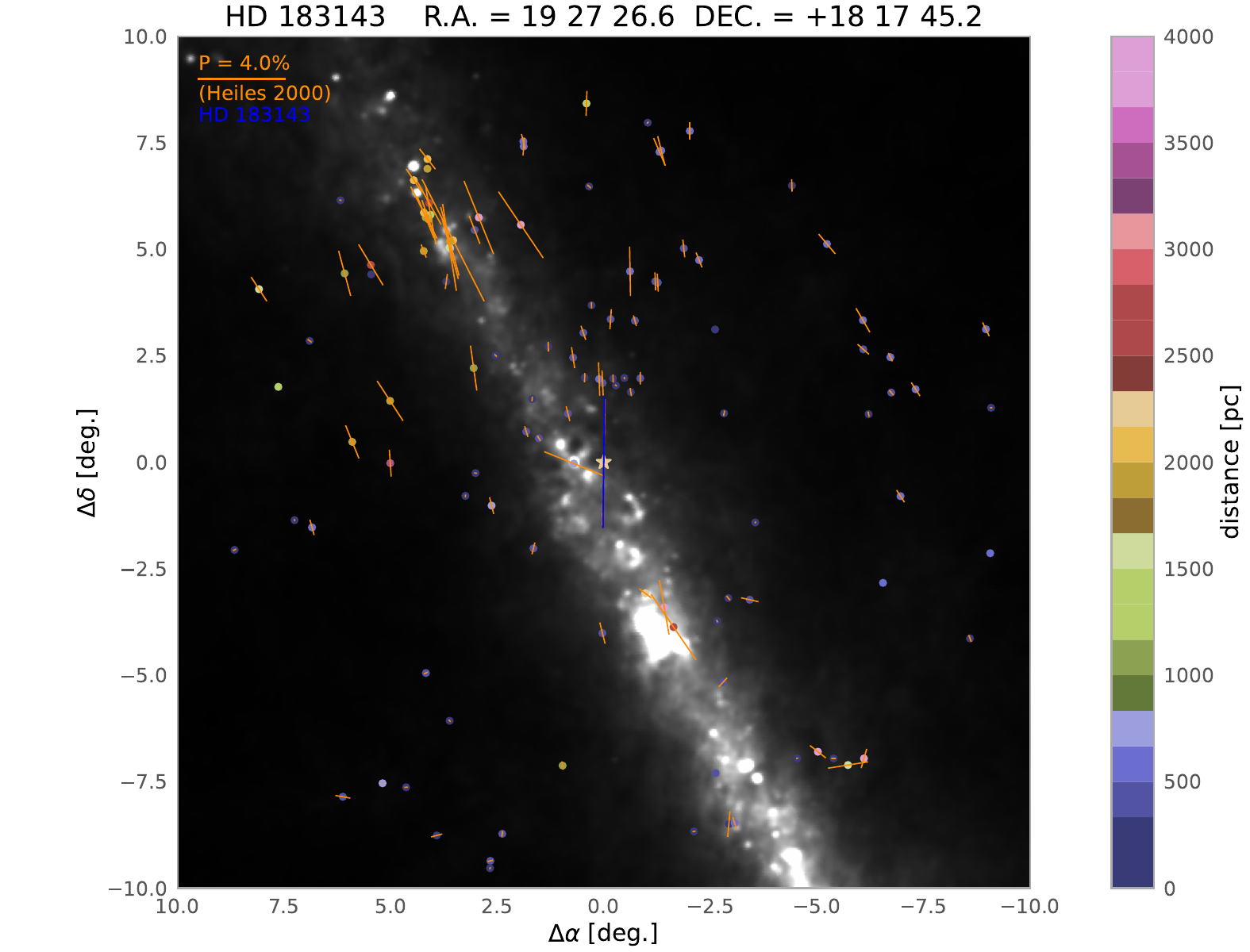}}\
    \subfloat{\includegraphics[width=0.33\textwidth, angle=0]{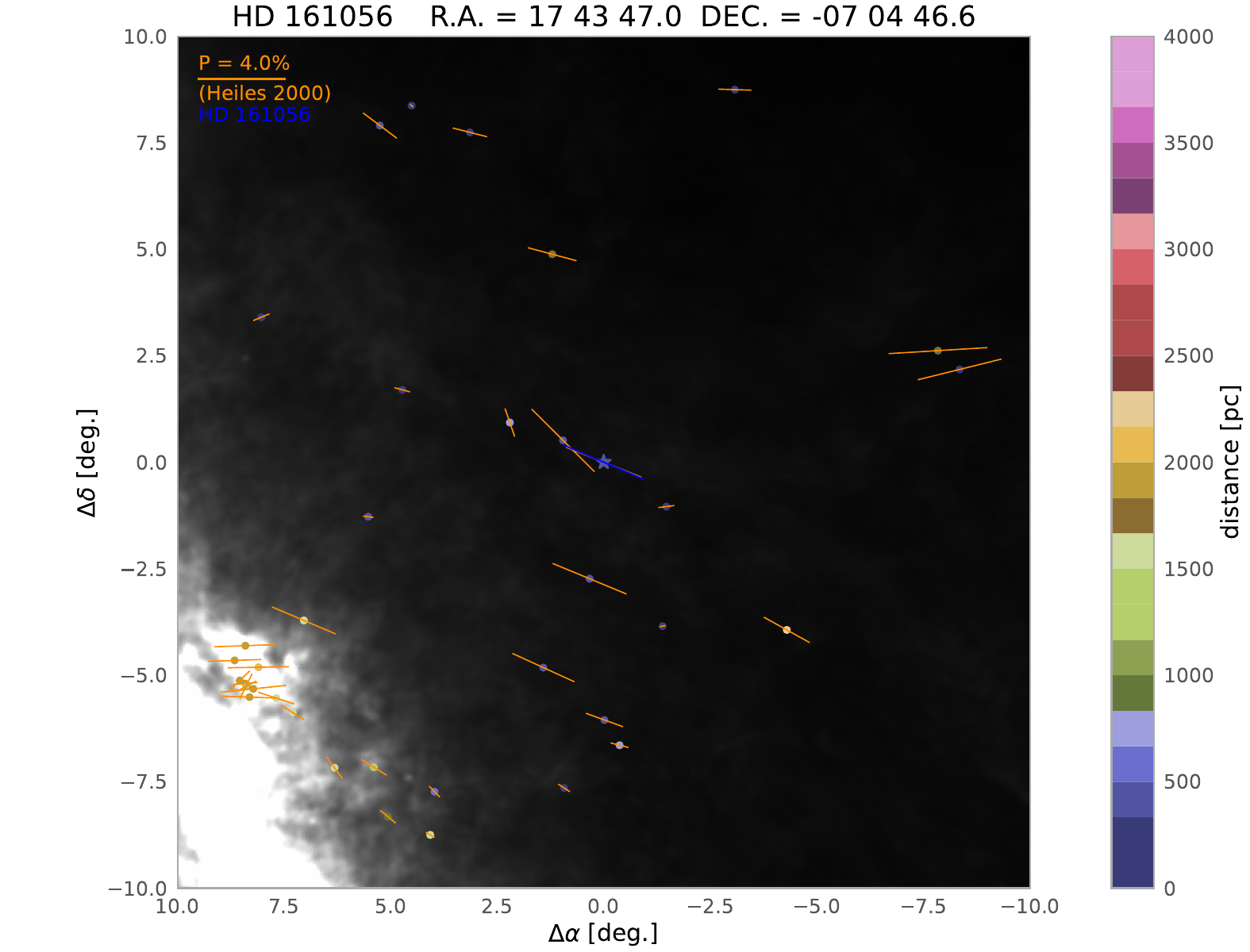}}\
    \subfloat{\includegraphics[width=0.33\textwidth, angle=0]{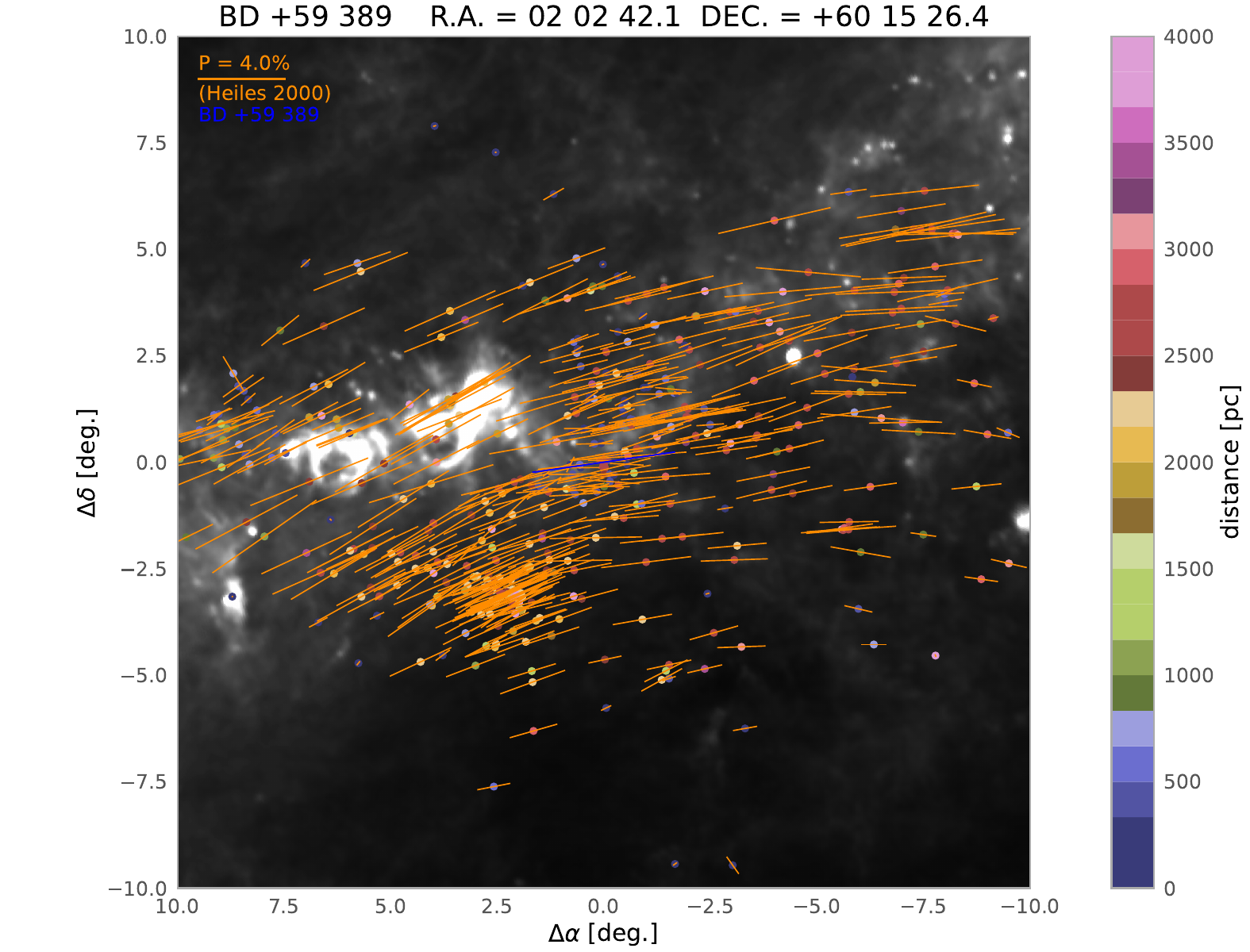}}\\
    \subfloat{\includegraphics[width=0.33\textwidth, angle=0]{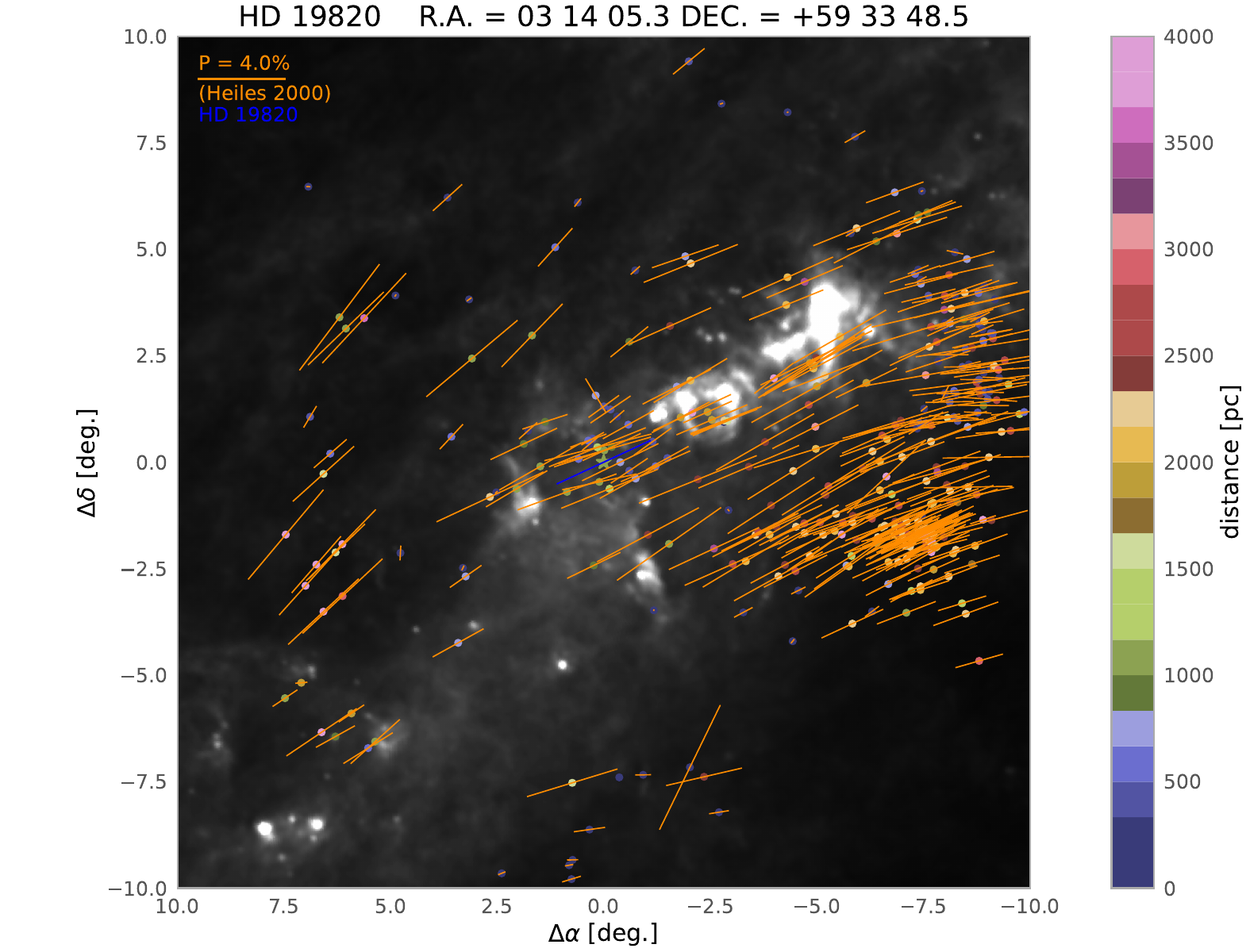}}\
    \subfloat{\includegraphics[width=0.33\textwidth, angle=0]{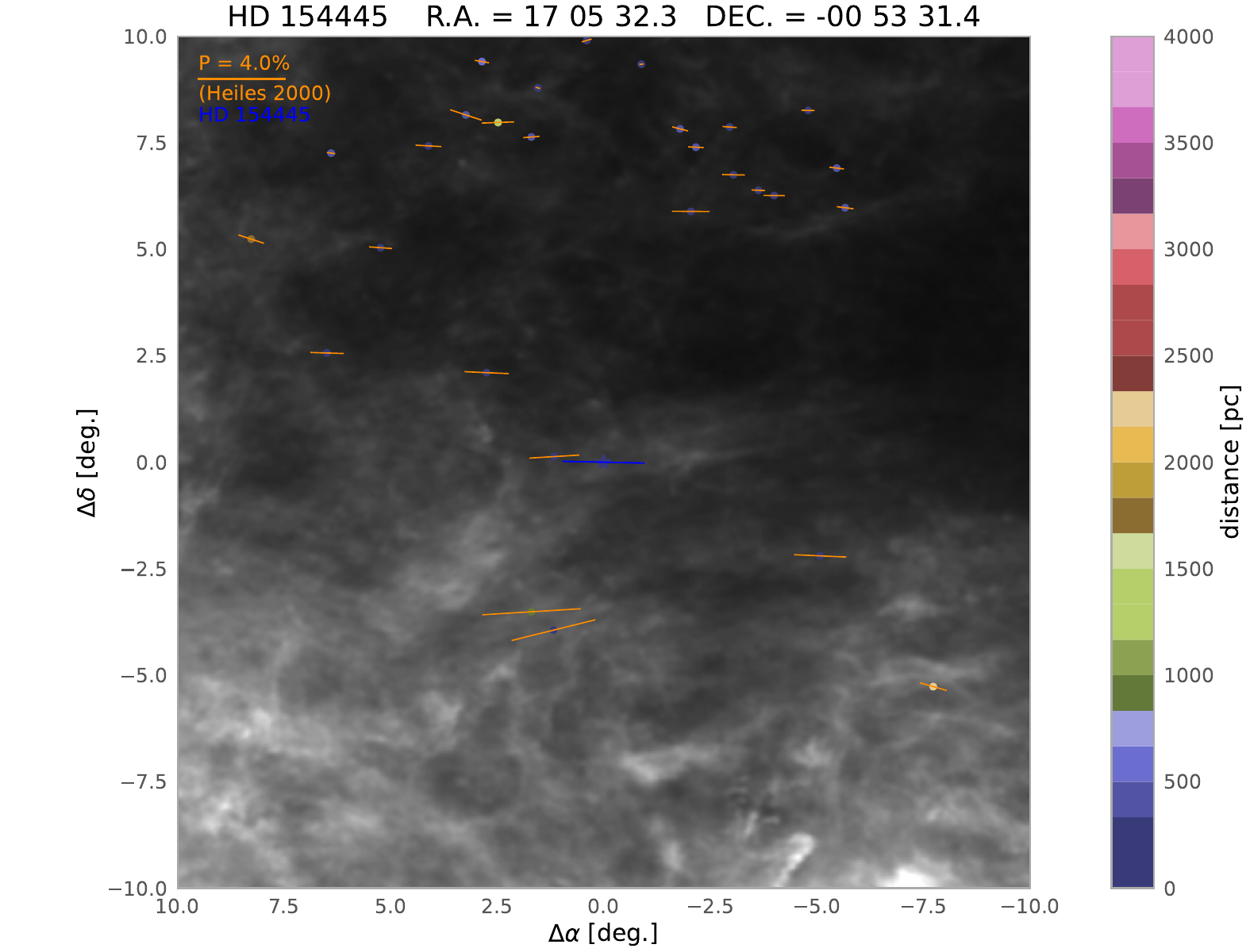}}\
    \subfloat{\includegraphics[width=0.33\textwidth, angle=0]{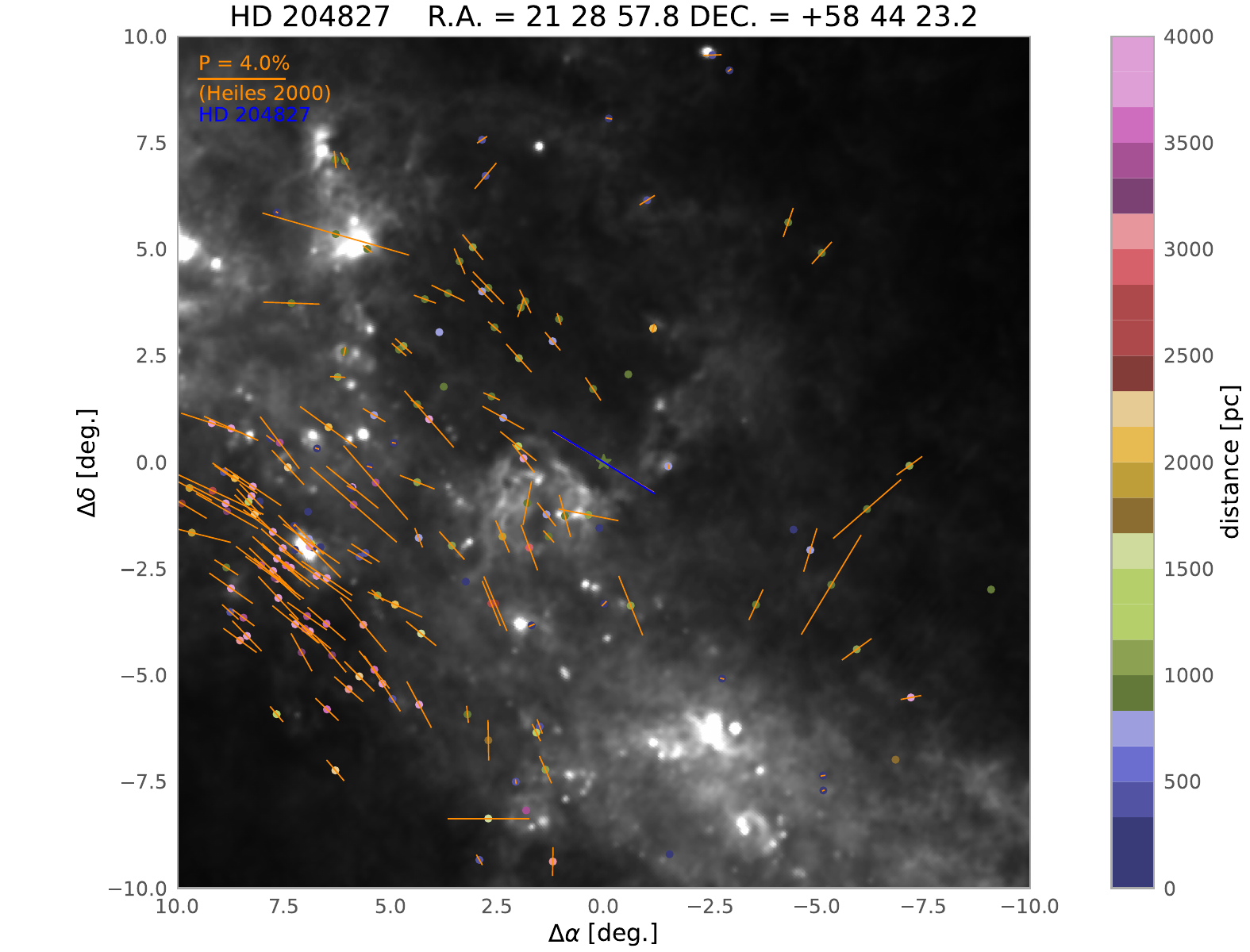}}\\
    \subfloat{\includegraphics[width=0.33\textwidth, angle=0]{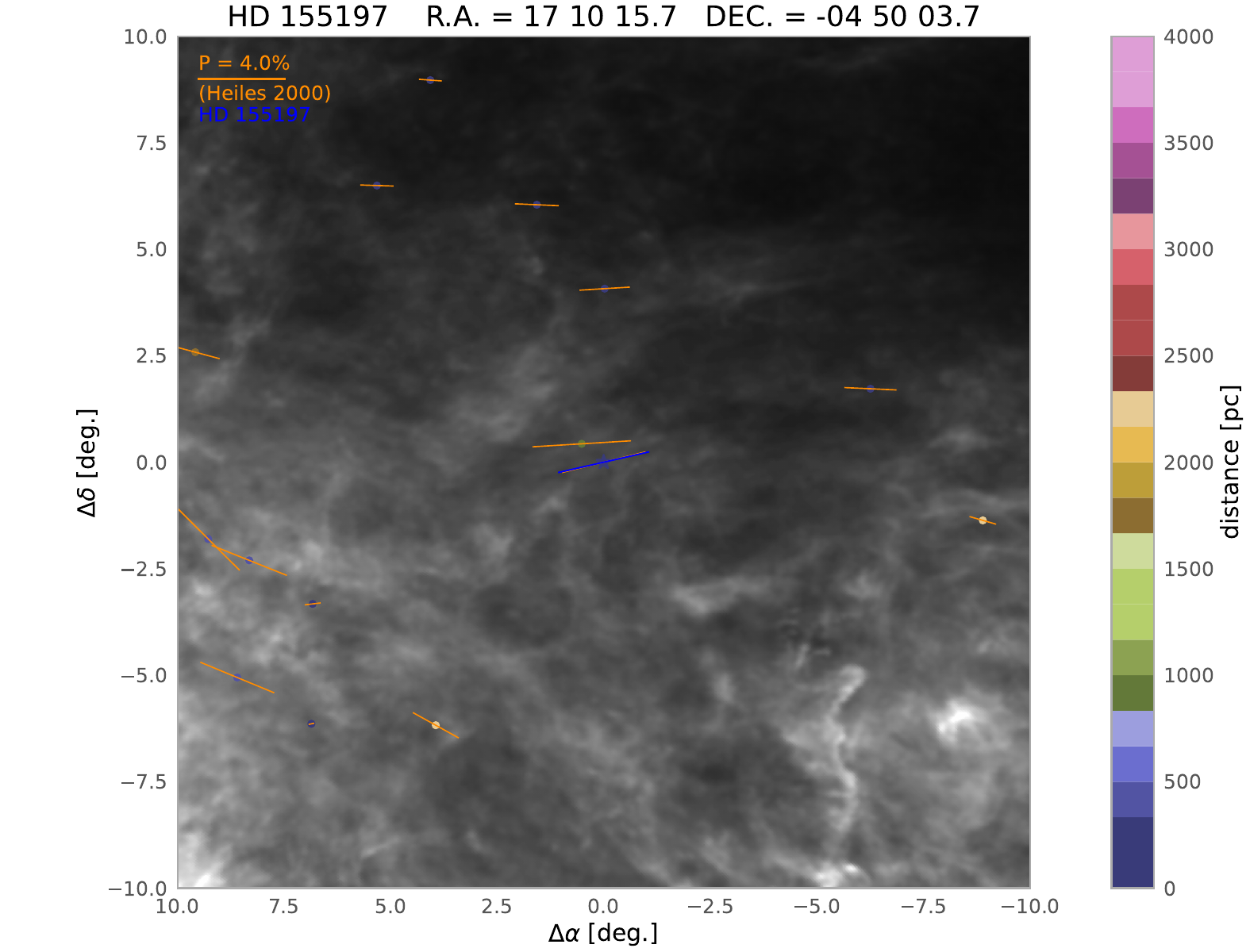}}    
    \caption{Interstellar polarisation of the field stars around observed high-polarisation standards \citep{2000AJ....119..923H}. High-polarisation standards are marked with a star symbol and are located at the centre of the images. The degree of polarisation is proportional to the length of the orange straight lines. The horizontal bar at the top left presents 4.0\% polarisation. The colour of each star corresponds to its distance. The distance is based on Gaia Early Data Release 3 \citep{2021AJ....161..147B}. The background image represents 100~$\mu$m dust emission maps \citep{1998ApJ...500..525S}.}
    \label{3D_pol}
\end{figure*}

To demonstrate how the PA of highly polarised standard stars is oriented, we created 3D maps. In Fig.~\ref{3D_pol}, we present 3D polarisation maps in the vicinity of the observed high-polarisation standards. We used data from the \citet{2000AJ....119..923H} catalogues. The colour of each star corresponds to its distance, based on data from Gaia Early Data Release 3 \citep{2021AJ....161..147B}. The distance to BD +25 727 is unclear that is why it is presented with a blue dot in Fig.~\ref{3D_pol}. High-polarisation standards are marked with a star symbol and are located at the centre of the images. The blue line over the high-polarisation standards symbol represents the catalogue values of the PA and the degree of polarisation itself as presented in Table \ref{tabl.list.sts}.  
In most cases, the PA of the observed high-polarised standard star is aligned with the PAs of neighbouring stars. An exception to this is observed in HD 183143. The background image represents 100 $\mu$m dust emission maps \citep{1998ApJ...500..525S}.

\end{appendix}
\end{document}